\documentclass[letterpaper,twocolumn,10pt]{article}
\usepackage{zhanggroup}
\usepackage{caption,subcaption}
\usepackage{tikz}
\usepackage{amsfonts}
\usepackage{xspace} 
\usepackage{mathrsfs}
\usepackage{subcaption}
\usepackage{booktabs}
\usepackage{multirow}
\usepackage{tcolorbox}
\usepackage[toc,page]{appendix}
\usepackage{etoolbox}
\usepackage[absolute]{textpos}

\newcommand{\mypara}[1]{\smallskip\noindent{\bf {#1}.}}

\begin{document}

\begin{textblock}{13}(1.5,1)
\centering
To Appear in the 16th International Conference on Web and Social Media (ICWSM), 2022.
\end{textblock}

\title{\Large \bf On Xing Tian and the Perseverance of Anti-China Sentiment Online}

\date{}

\author{
    Xinyue Shen,\textsuperscript{\rm 1}
    Xinlei He,\textsuperscript{\rm 1}
    Michael Backes,\textsuperscript{\rm 1}\\
    Jeremy Blackburn,\textsuperscript{\rm 2}
    Savvas Zannettou,\textsuperscript{\rm 3}
    Yang Zhang\textsuperscript{\rm 1}
    \\
    \\
    \textsuperscript{\rm 1}CISPA Helmholtz Center for Information Security \ \ \ 
    \textsuperscript{\rm 2}Binghamton University \ \ \ 
    \textsuperscript{\rm 3}TU Delft
}

\maketitle

\begin{abstract}

Sinophobia, anti-Chinese sentiment, has existed on the Web for a long time.
The outbreak of COVID-19 and the extended quarantine has further amplified it.
However, we lack a quantitative understanding of the cause of Sinophobia as well as how it evolves over time.  
In this paper, we conduct a large-scale longitudinal measurement of Sinophobia, between 2016 and 2021, on two mainstream and fringe Web communities.
By analyzing 8B posts from Reddit and 206M posts from 4chan's /pol/, we investigate the origins, evolution, and content of Sinophobia.
We find that, anti-Chinese content may be evoked by political events not directly related to China, e.g., the U.S. withdrawal from the Paris Agreement.
And during the COVID-19 pandemic, daily usage of Sinophobic slurs has significantly increased even with the hate-speech ban policy.
We also show that the semantic meaning of the words ``China'' and ``Chinese'' are shifting towards Sinophobic slurs with the rise of COVID-19 and remain the same in the pandemic period.
We further use topic modeling to show the topics of Sinophobic discussion are pretty diverse and broad.
We find that both Web communities share some common Sinophobic topics like ethnics, economics and commerce, weapons and military, foreign relations, etc.
However, compared to 4chan's /pol/, more daily life-related topics including food, game, and stock are found in Reddit.
Our finding also reveals that the topics related to COVID-19 and blaming the Chinese government are more prevalent in the pandemic period.
To the best of our knowledge, this paper is the longest quantitative measurement of Sinophobia.

\end{abstract}

\section{Introduction}

The story of Xingtian speaks of a deity that fought against the Supreme Divinity.
Although Xingtian's army lost, he refused to stop fighting, thus extreme actions were taken: he was decapitated and his head buried under a mountain.
However, this was still not enough to stop Xingtian.
He continued to fight, using his nipples to see and his bellybutton to speak.
The story of Xingtian is one of persistence, and has parallels to worrying behavior on the Web: it continues to persist.

While Sinophobia, i.e., anti-Chinese sentiment, had a staggering rise after the COVID-19 pandemic began, it has persisted for hundreds of years.
For example, in 1882 the Chinese Exclusion Act was passed, which barred Chinese workers from entering the US until its repeal 60 years later in 1943~\cite{AMERICANSIN1882}.
In 2013, a survey conducted by the Pew Research Center~\cite{KWSSPBPBDSBGH13} showed that Sinophobia persisted in the West, e.g., only 34\% of Americans, 28\% of Italians, and 28\% of Germans have a favorable opinion of China.

With the developments of the Internet, mediums like texts, images, and videos are created and shared at ever increasing volume.
However, there is also a downside of the Web, e.g., the rise of fringe communities such as 4chan's Politically Incorrect board (/pol/).

Sinophobia is indeed a ``popular'' topic discussed on fringe Web communities~\cite{ASIANHATE, ASIAHATE2,TSLBSZZ21}.
The effects of Sinophobia are not only seen on the Web, but also in the physical world.
Relia et al.~\cite{RLCC19} provide evidence that online racist activity correlates with hate crimes.
The outbreak of COVID-19 has amplified Sinophobia on fringe Web communities and mainstream social media like Twitter and Reddit~\cite{TSLBSZZ21}.

With the advent of the COVID-19 pandemic and its origins in China, Sinophobia has become a topic of research.
Tahmasbi et al.~\cite{TSLBSZZ21} study the raise of Sinophobic behaviors after COVID-19 on both fringe and mainstream Web communities over 5 months and Ziems et al.~\cite{ZHSK21} investigate the evolution of anti-asian on Twitter across three months after the outbreak of COVID-19.
However, to the best of our knowledge, there is no study that analyzes Sinophobia over a longer period of time (i.e., \emph{before} the COVID-19 era), and thus there is a meaningful gap in our understanding of the evolution of Sinophobia on social media.
Also, the previous studies~\cite{TSLBSZZ21} only focus on word-level analysis, a more comprehensive content-level analysis with respect to toxicity and topics is missing.
More importantly, as the world begins to recover from COVID-19, it remains unclear as to whether Sinophobic behavior has seen a downtrend as well.

In this paper, we perform a large-scale longitudinal measurement of how Sinophobia has ebbed and flowed from 2016 to 2021 over two Web communities: 4chan's Politically Incorrect board (/pol/) and Reddit.
With over 206,329,303 posts on /pol/ and 8,118,465,218 posts on Reddit, we quantify Sinophobic behaviors with regard to its origins, evolution, and content.

Concretely, we first measure the temporal patterns of China-related posts and dive into the detail of Sinophobic slurs. 
We find that Sinophobia was prevalent before COVID-19 and was correlated to political events such as the Trump–Tsai Call~\cite{TrumpTsaicall} and Hong Kong Protests~\cite{HKPROTESTS}. 
More importantly, many of such political events are not directly related to China, e.g.,  the U.S. withdrawal from the Paris Agreement~\cite{USWITHDRAWPARIS} and inauguration of Joe Biden~\cite{BIDENINAUGURATION}.
COVID-19, however, saw a drastic change in Sinophobia; the daily usage of Sinophobic slurs increases substantially after COVID-19, e.g., the utilization frequency of Sinophobic word ``chicoms'' increases $5.2\times$.
Even with a hate-speech ban policy~\cite{Reddit_ban_hate}, slurs still sneak on Reddit, which calls for the mainstream community to take more actions and responsibility.
Using Google's Perspective API, we find that posts related to China and Chinese are more toxic than a baseline set of comments, and this drastically increases during the pandemic period (see the section ``Temporal Analysis'' for more details).

We then analyze the content of posts and find that ``china'' and ``chinese'' have had a sharp shift away from referring to the country/Chinese government and towards Sinophobic slurs.
For instance, on /pol/, the meaning of ``china'' is close to ``taiwan'' and ``asia'' in 2016, while shifting to ``chink'' and ``chinkland'' in 2020.
By performing topic extraction, we find a diverse set of discussions.
For instance, /pol/ and Reddit both have Sinophobic topics that are related to ethnics, economics and commerce, weapons and military, foreign relations, etc.
Compared to /pol/, Sinophobic topics on Reddit are more diverse in languages and cover a wider range of topics that are related to daily life such as food, game, and stock.
We also observe that the Sinophobic topics switch after the outbreak of COVID-19.
For instance, people show more interest in pandemic-related topics, with an increase of $11\times$ on 4chan's /pol/ (topic 3) and $5.32\times$on Reddit (topic 7).
our analysis also reveals that, in the pandemic period, users express more toxic posts towards the Chinese government with anger, e.g., the scale of such topics exploded to $1.25\times$ on 4chan's /pol/ (topic 5) and $1.71\times$ on Reddit (topic 12).

\mypara{Disclaimer}
Note that content posted on both Web communities we study can be considered racist and offensive. 
In the rest of this paper, we do not censor any language to better illustrate the peculiarities of the problem.
We inform the reader that this paper contains content that is likely to be offensive and disturbing.

\section{Related Work}

We now review relevant previous work.
We report on two areas:
1) study on Sinophobia across years;
2) measurement of racial activity on social networks.

\mypara{Sinophobia studies across years}
East Asian prejudice has remained for a long time in the West: 
in the 19th century, the racial phrase ``yellow peril'' is created to insult Chinese and has been proved re-visited when COVID-19 outbreaks~\cite{LJ21}.
Several surveys and studies depicted the unfavorable attitudes towards China prior to COVID-19~\cite{B14, SINSUVERY2017,SINSURVEY2019}.
Tahmasbi et al.~\cite{TSLBSZZ21} firstly examined Sinophobia during the outbreak of COVID-19 on Twitter and 4chan's /pol/.
They mainly focused on Sinophobic behaviors in word-level and find that COVID-19 indeed provokes the emergence of Sinophobic slurs.
However, their dataset only acrosses 5 months and therefore lacks a longitudinal perspective.
Nguyen et al.~\cite{NCDHKHPNYGACGN20} conducted sentiment analysis on race-related tweets and find the proportion of negative tweets referencing Asians increased by 68.4\% till March 2020.

\mypara{Racism analysis on the internet}
Besides Sinophobic behaviors, several papers measure online racial activities.
Zannettou et al.~\cite{ZFBB20} presented a quantitative analysis on online antisemitism.
Cervi~\cite{C20} and Chandra et al.~\cite{CRSGBK21} analyzed Islamophobia.
Mittos et al.~\cite{MZBC20} focused on ethnic discrimination on Reddit and 4chan's /pol/ via understanding genetic testing conversations. 
Yang et al.~\cite{YC18} investigated on self-narration of racial discrimination in Reddit threads.

\mypara{Remarks}
To the best of our knowledge, a quantitative understanding of the cause of Sinophobia as well as how it evolves over several years is lacking.
In this paper, we aim to bridge the gap.

\begin{table*}
\centering
\small
{
  \begin{tabular}{c|c|c|c}
    \toprule
    Dataset& \#Posts & \#Filtered Posts & Time Range\\
    \midrule
    4chan & 206,329,303  & 2,193,410 & 2016.06.30-2021.03.18\\
    Reddit & 8,118,465,218 & 26,183,882 & 2016.06.01-2021.03.31 \\
  \bottomrule
\end{tabular}
}
\caption{Overview of the 4chan and Reddit datasets. Filtered posts refer to posts containing the term ``china'' or ``chinese''.}
\label{table:dataset}
\end{table*}

\section{Datasets}

In this section, we present our datasets from 4chan's Politically Incorrect board (/pol/) and Reddit. 
Table~\ref{table:dataset} summarizes the two collected datasets.

\mypara{4chan's /pol/}
4chan is an anonymous imageboard organized by sub-communities named ``boards'', each driven by a specific topic of interest.
In this work, we focus on the Politically Incorrect board (/pol/) as it is the mainboard for the discussion of politics and world events.
Additionally, 4chan is ephemeral as it maintains a limited number of active threads and permanently deletes threads after a week. 
We choose to include 4chan in our analysis as it is a fringe Web community, known for the dissemination of toxic and offensive language~\cite{HOCKLSSB17}, hence it is likely to include a considerable amount of Sinophobic language~\cite{TSLBSZZ21}.
We obtain the data via the official API provided by 4chan~\cite{4CHANAPI} and follow the data collection approach described in~\cite{PZCSB20}.
In total, we collect 206,329,303 posts between June 30, 2016 to March 18, 2021.

\mypara{Reddit}
Reddit is a social news aggregation and discussion website organized by subreddits, which are sub-communities created by users. 
Registered members are allowed to submit content to these subreddits, get replies from other users, or receive upvotes or downvotes by other members (upvotes and downvotes determine the popularity of content within the platform).
We include Reddit in our analysis because of the platform's popularity~\cite{REDDITDAILY} and the platform's diversity (i.e., with hundreds of millions of subreddits the platforms cover a huge set of interests).
Due to the platform's popularity and diversity, we expect that discussions related to Chinese and Asian people in general are happening on Reddit, hence allowing us to include in our analysis a more mainstream platform (compared to 4chan).
In this paper, we collect all posts and comments on Reddit from~\cite{BZKSB20}. 
In total, we gather 8,118,465,218 posts on Reddit between June 1, 2016 and March 31, 2021.

\section{Temporal Analysis}\label{section:temporal_analysis}

\begin{figure*}[ht]
  \centering
	\begin{subfigure}{0.48\linewidth}
    	\includegraphics[width=\textwidth]{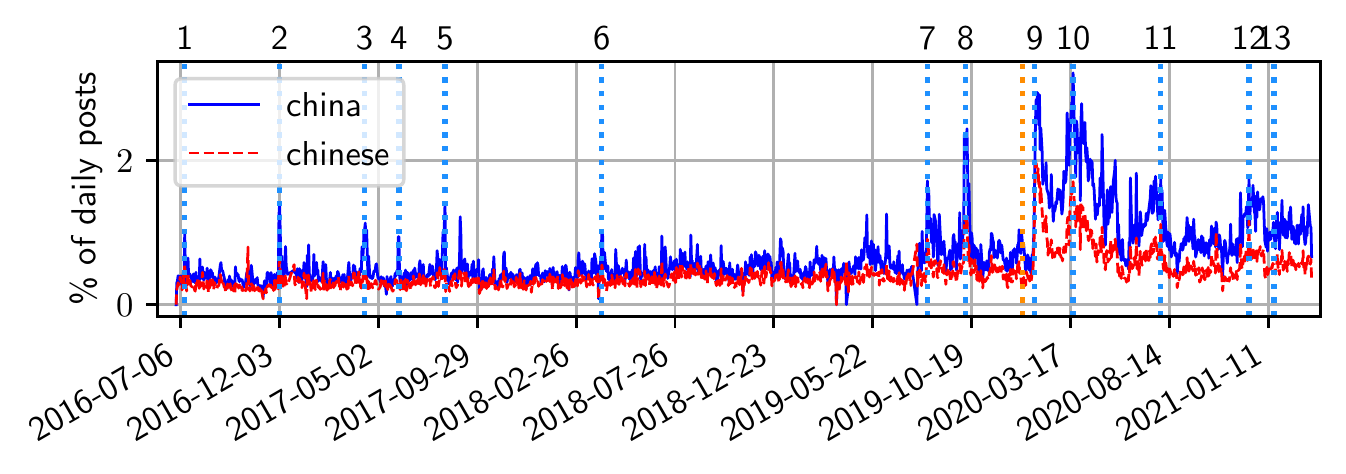}
    	\caption{Percentage of daily posts on 4chan's /pol/}
    \end{subfigure}\hfill
	\begin{subfigure}{0.48\linewidth}
		\includegraphics[width=\textwidth]{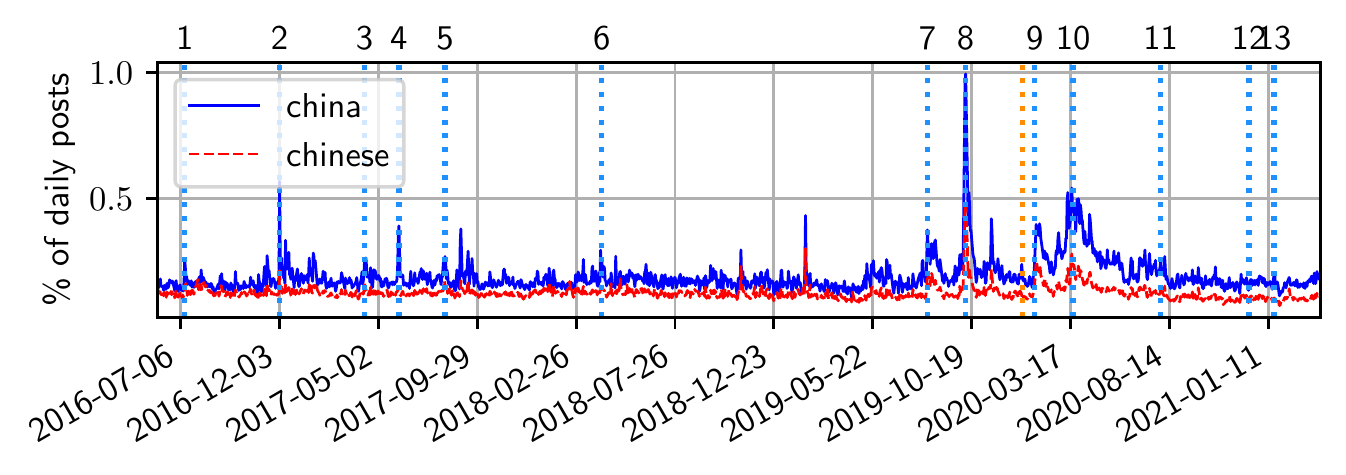}
		\caption{Percentage of daily posts on Reddit}
	\end{subfigure}\hfill

\caption{Percentage of daily posts that contains ``china'' or ``chinese'' on 4chan's /pol/ and Reddit. The orange dotted line is the split day (2020-01-04) of pre-pandemic and pandemic periods. The blue dotted lines refer to political events (see Table~\ref{table:events})}.
\label{figure:china_chinese_occ_per}
\end{figure*}

\begin{table}[t]
\centering
\small
{
\setlength\tabcolsep{3pt}
\begin{tabular}{p{0.05\linewidth}|p{0.2\linewidth}|p{0.65\linewidth}}
\toprule
\textbf{NO.} & \textbf{Peak Day} & \textbf{Event}                                                          \\
\midrule
1 &
  2016-07-12 &
Territorial disputes in the South China Sea~\cite{EVENT_SOUTHSEA} \\
2            & 2016-12-03             & Trump–Tsai call~\cite{TrumpTsaicall}                                                \\
3 &
  2017-04-11 &
  N.K. announces it will nuke U.S. at first sign of pre-emptive strike.~\cite{NORTHNUKEFIRST} \\
4            & 2017-06-02             & United States withdraw from the Paris Agreement~\cite{USWITHDRAWPARIS}              \\
5            & 2017-08-11             & China pledges neutrality - unless US strikes North Korea first~\cite{CHINAHELPNK} \\
6            & 2018-04-06             & China hits back after Trump threatens \$100bn in tariffs~\cite{CHINATRADEWAR_HITBACK}       \\
7            & 2019-08-13             & Police brutality protests in HongKong Protests~\cite{HKPROTEST_AUG}                 \\
8            & 2019-10-10             & Hongkong protests~\cite{HKPROTEST_OCT}                                              \\
9            & 2020-01-23             & COVID-19 lockdown in China~\cite{COVID19LOCKDOWN}                                     \\
10           & 2020-03-21             & COVID-19 spread worldwide~\cite{COVID19SPREAD}                                      \\
11           & 2020-08-01             & Trump said he will ban TikTok in the US~\cite{TICKTOKBAN_AUG1}                        \\
12 &
  2020-12-13 &
  A major leak containing a register with the details of nearly two million CCP members has occurred and exposed to the public~\cite{CCPLEAK} \\
13           & 2021-01-20             & Inauguration of Joe Biden ~\cite{BIDENINAUGURATION}                      \\
\bottomrule
\end{tabular}
}
\caption{Major Events, annotated on Figure~\ref{figure:china_chinese_occ_per}.}
\label{table:events}
\end{table}

To identify the scope of Sinophobia, we start with an investigation on the temporal patterns of the term ``china'' and ``chinese.''
We then elaborate on Sinophobic slur findings and their influence as well as implications.
Next, we study the meaning behind the changes in terms of correlation and perspectives.

\subsection{Trend Analysis}

\label{section: trend_analysis}
We first focus on the daily usage of ``china'' and ``chinese'' on 4chan's /pol/ and Reddit.
To do this, for each post, we convert the text to lowercase, perform tokenization using NLTK~\cite{BKL09}, and  search for the terms ``china'' and ``chinese.''

Figure~\ref{figure:china_chinese_occ_per} shows the daily percentage (over all posts per day) of the terms ``china'' and ``chinese'' on 4chan's /pol/ and Reddit. 
We annotate the figure with the day that the World Health Organization (WHO) first tweeted about coronavirus (orange dotted line, January 4, 2020)~\cite{WHOCOVIDTWEET}, which is the first day that COVID-19 has officially entered the global spotlight on Web communities. 
In this paper, we consider the period before January 4, 2020 as the \textit{pre-pandemic period} and the period after January 4, 2020, as the \textit{pandemic period}.
We also detect 13 peaks in our datasets and annotate them with blue dotted lines by using peak detection~\cite{scipy_peak_detection}.
By manually checking around 2,000 posts in each peak, we find discussion on each peak corresponds to at least one event that happened around that day (see Table~\ref{table:events}).
Surprisingly, we find that events not directly related to China may also evoke discussions related to China and Chinese people.
For instance, on June 1, 2017, President Trump announced that the U.S. would end all participation in the 2015 Paris Agreement on climate change mitigation~\cite{USWITHDRAWPARIS}. 
This announcement evokes numerous posts of China and Chinese on both 4chan's /pol/ and Reddit and reaches its peak on June 2, 2017.
At this peak, China and Chinese are frequently mentioned for their roles playing in environmental protection.
For instance, a Reddit user posts:`` \textit{fucking china is the one producing the co2 bigle.''}

We then measure the peak width of each event via~\cite{scipy_peak_detection} and regard it as the interest period of each event. 
The relative height at which the peak width is measured is 0.5.
We find that the pattern of interest period thoroughly changes during the pandemic period.
For instance, in the pre-pandemic period, the average interest time of ``china'' (``chinese'') remains 2.76 (2.83) days.
However, it increases to 8.00 (9.28) days in the pandemic period, indicating prolonged discussions and extended interest around China and Chinese people.

The daily usage of ``china'' and ``chinese'' also rises in the pandemic period.
Take 4chan's /pol/ as an example, the average percentage of posts per day containing ``china'' (``chinese'') mounts from 0.50\% (0.34\%) to 1.23\% (0.66\%) after the outbreak of COVID-19. 
For Reddit, it also increases from 0.18\% to 0.22\%.
This finding implies that COVID-19 indeed raise the attention towards China and Chinese people on fringe and mainstream Web communities.

\begin{figure*}[!t]
  \centering
  	\begin{subfigure}{.33\linewidth}
		\centering
		\includegraphics[width= \linewidth]{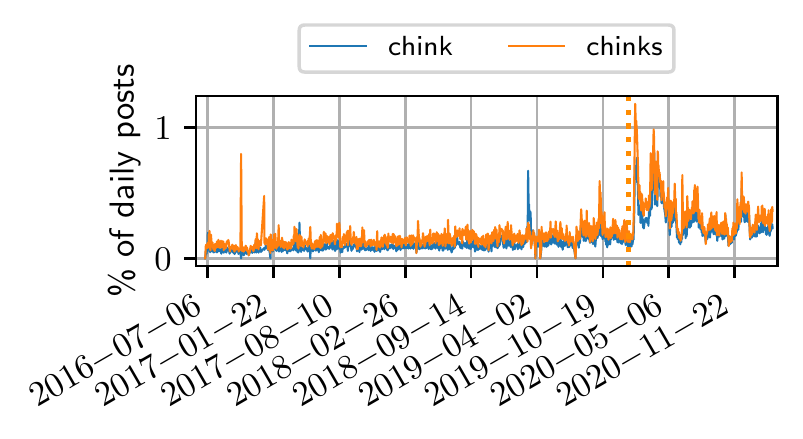}
	\end{subfigure}\hfill
	\begin{subfigure}{.33\linewidth}
    	\centering 
    	\includegraphics[width=\textwidth]{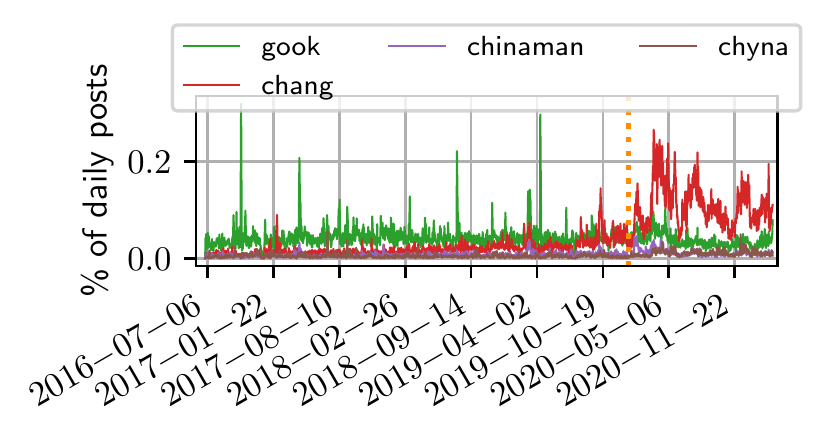}
    \end{subfigure}\hfill
	\begin{subfigure}{.33\linewidth}
    	\centering 
    	\includegraphics[width=\textwidth]{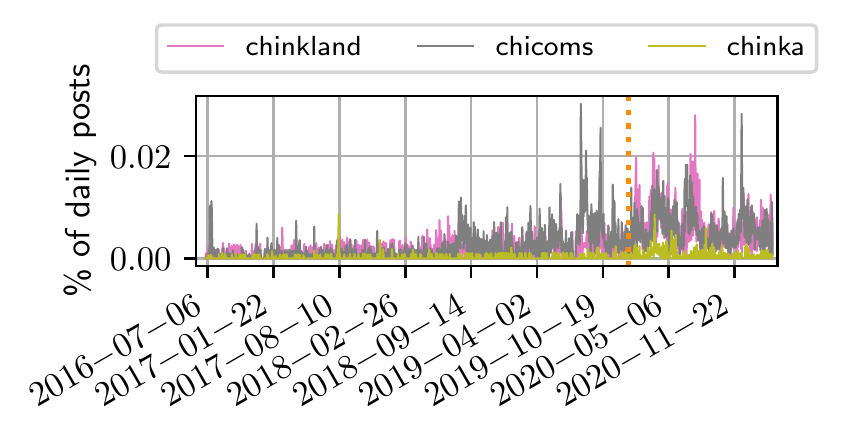}
    \end{subfigure}\hfill
\caption{Percentage of posts that contains Sinophobic slurs on 4chan's /pol/. The orange line refers to the split day (Jan 4, 2020).}
\label{figure:slur_pol}
\end{figure*}

\begin{figure*}[!t]
  \centering
  	\begin{subfigure}{.33\linewidth}
		\centering
		\includegraphics[width= \linewidth]{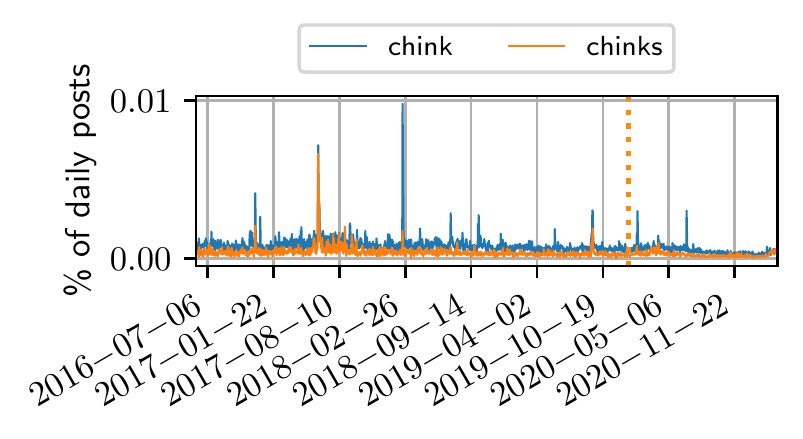}
	\end{subfigure}\hfill
	\begin{subfigure}{.33\linewidth}
    	\centering 
    	\includegraphics[width=\textwidth]{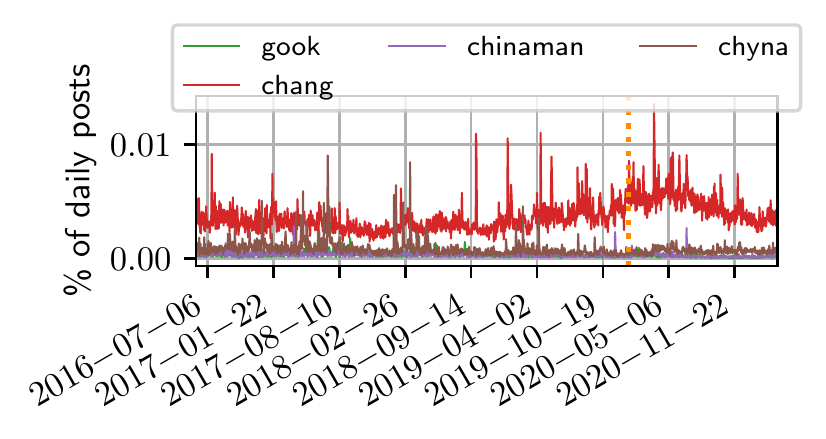}
    \end{subfigure}\hfill
	\begin{subfigure}{.33\linewidth}
    	\centering 
    	\includegraphics[width=\textwidth]{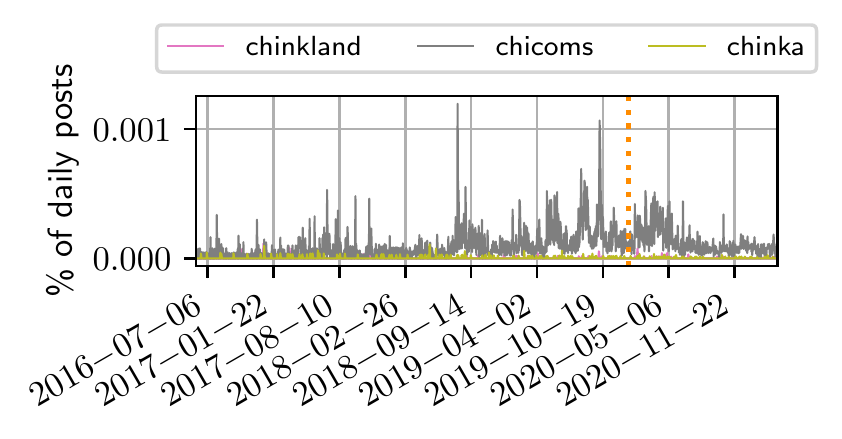}
    \end{subfigure}\hfill
\caption{Percentage of posts that contain Sinophobic slurs on Reddit. The orange line refers to the split day (Jan 4, 2020).}
\label{figure: slur_reddit}
\end{figure*}

\begin{table*}[!t]
\centering
\small{
\setlength\tabcolsep{5pt}
\begin{tabular}{lr|lr|lr|lr|lr|lr}
\toprule
\multicolumn{6}{c|}{\textbf{4chan's /pol/}}                                             & \multicolumn{6}{c}{\textbf{Reddit}}                         \\
\midrule
\multicolumn{2}{c|}{\textbf{china}} &
  \multicolumn{2}{c|}{\textbf{chinese}} &
  \multicolumn{2}{c|}{\textbf{china/chinese + hate}} &
  \multicolumn{2}{c|}{\textbf{china}} &
  \multicolumn{2}{c|}{\textbf{chinese}} &
  \multicolumn{2}{c}{\textbf{china/chinese + hate}} \\
   \midrule
chinas                 & 0.786    & chine                   & 0.746     & \textbf{chink}                & 0.670            & taiwan            & 0.761         & taiwanese           & 0.757         & ccp                           & 0.559            \\
\textbf{chyna}         & 0.756    & chineese                & 0.732     & \textbf{chyna}                & 0.599            & india             & 0.746         & china               & 0.730         & chinese                       & 0.575             \\
prc                    & 0.730    & \textbf{chink}          & 0.719     & \textbf{chinaman}             & 0.612            & chinese           & 0.730         & korean              & 0.664         & russia                        & 0.558            \\
chine                  & 0.725    & china                   & 0.718     & chinas                        & 0.571            & russia            & 0.720         & vietnamese          & 0.656          & chinas                        & 0.547            \\
chinese                & 0.718    & ccp                     & 0.684     & chine                         & 0.563            & chinas            & 0.713         & indonesian          & 0.647         & indian                        & 0.539             \\
taiwan                 & 0.695    & \textbf{chinaman}       & 0.675     & chineese                      & 0.574            & mainland          & 0.709         & indian              & 0.646         & india                         & 0.527             \\
ccp                    & 0.693    & prc                     & 0.664     & \textbf{chinkland}            & 0.570            & beijing           & 0.708         & japanese            & 0.643         & \textbf{chinaman}             & 0.543             \\
india                  & 0.681    & chiense                 & 0.662     & \textbf{gook}                 & 0.606            & prc               & 0.697         & westerner           & 0.613         & taiwanese                     & 0.536            \\
\textbf{chinkland}     & 0.677    & taiwanese               & 0.649     & \textbf{chicoms}              & 0.549            & ccp               & 0.679          & mainland            & 0.610         & china                         & 0.526            \\
nk                     & 0.677    & mainland                & 0.628     & ccp                           & 0.572            & country           & 0.678         & chineese            & 0.609         & vietnamese                    & 0.537            \\
japan                  & 0.653    & chinse                  & 0.625     & \textbf{chang}                & 0.566            & iran              & 0.663         & taiwan              & 0.599         & \textbf{chink}                & 0.532            \\
russia                 & 0.651    & mainlanders             & 0.624     & \textbf{chinks}               & 0.592            & korea             & 0.662         & foreigner           & 0.598          & commie                        & 0.567            \\
mainland               & 0.643    & \textbf{chinkland}      & 0.617     & mainlander                    & 0.566            & pakistan          & 0.630         & beijing             & 0.598         & westerner                     & 0.531            \\
usa                    & 0.633    & japanese                & 0.611     & \textbf{chinka}               & 0.557            & dprk              & 0.630         & laowai              & 0.585         & laowai                        & 0.528            \\
\textbf{chicoms}       & 0.628    & \textbf{chinks}         & 0.605     & chinese                       & 0.632            & europe            & 0.630         & malaysian           & 0.583         & \textbf{gook}                 & 0.567     \\      
\bottomrule
\end{tabular}
}
\caption{Top 15 most similar words to the terms ``china''/``chinese'', and top 15 most frequently occurring words towards combined hateful weight vectors. The value denotes the cosine similarity.}
\label{table:word2vec_slur}
\end{table*}

\subsection{Racial Slurs} 
\label{section: racial_slurs}

We then take a further step to understand the origin, usage, and evolution of Sinophobic slurs.
Concretely, we semi-automatically capture anti-China and anti-Chinese slur words via word2vec models~\cite{MSCCD13} trained on the entire 4chan's /pol/ dataset and a dataset of 1\% randomly sampled posts from Reddit. 
Before training the models, we first preprocess the two datasets as follows: 1) we convert the content of posts to lowercase and expand contractions such as ``it's'' to ``it is''; 2) we remove punctuation, stopwords, HTML tags, and numbers.
Next, we train word2vec models for each Web community on the whole pre-processed corpus with all words that appear at least 20 times.
We set the context window to 5 following~\cite{GENSIM.Word2vec}. 
To semi-automatically identify hate words towards ``china'' and ``chinese'', we leverage 15 hate words following~\cite{HOCKLSSB17}, which are ``nigger,'' ``faggot,'' ``retard,'' ``bitch,'' ``idiot,'' ``cunt,'' ``kike,'' ``fag,'' ``nazi,'' ``trash,'' ``pussy,'' ``goy,'' ``frog,'' ``spic,'' and ``chink.'' 
Then, we perform operations (i.e., additions) on the word embeddings of each of the above hate words and the terms ``china''/``chinese'' and extract the 10 most similar words based on the resulting embeddings (e.g., extracting the most similar words to the embedding that is calculated from the addition of the embeddings for the words ``nigger'' and ``china''). 
By iteratively doing the same procedure for all combinations of words, we obtain 300 words.
We then count the number of appearances of each similar word and rank them.

In Table~\ref{table:word2vec_slur}, we report top15 most similar words to ``china'' and ``chinese,'' as well as the top 15  most frequently occurring words by combining the embeddings from the hate words and the terms ``china'' and ``chinese.''
By manually looking into them, we find nine derogatory terms referring to China and Chinese, including ``chink'' (slur word referring to Chinese and East Asian people)~\cite{CHINK}, ``chinks'' (plural of chink), ``chinkland'' (an offensive word referring to the land of chinks)~\cite{CHINKLAND}, ``chinka'' (derives from ``chink'' and the last syllable of the word ``nigga'')~\cite{CHINKA},  ``chinaman'' (evolved from its use in pejorative contexts regarding Chinese)~\cite{CHINAMAN}, ``chicoms'' (a contemptuous term used to refer to a Communist Chinese)~\cite{CHICOMS}, ``chang'' (an ethnic slur to Chinese, evolved from the Chinese language mocker)~\cite{CHANG}, ``chyna'' (an deliberately misspelled word to insult China)~\cite{CHYNA}, and ``gook'' (A derragatory term used against Asian  )\cite{GOOK}. 
An example post from 4chan's /pol/: \textit{``i can't believe there's actual people who side with fucking chyna. like seriously they are another level of shit people''}.
Another /pol/ user posts: \textit{``everyone is seriously making me hate chinks. we need to ship them all back to chinka''}.

Figure~\ref{figure:slur_pol} shows the daily proportion of Sinophobic slur words on 4chan's /pol/. 
We first observe that the usage of Sinophobic slurs coincides with political events.
For instance, on Aug 1, 2020, Trump announced he would ban TikTok~\cite{TICKTOKBAN_AUG1}, there is a sudden spike of all nine slurs on both Web communities.
In addition, the average daily usage of slurs also surges in the pandemic period.
For example, the utilization frequency of ``chicoms'' climbs $5.2\times$ in the pandemic period.
Note that this rise in slurs can not be simply attributed to the increase of ``china'' and ``chinese''.
Take ``chink'' and  ``chinks'' as an example.
The daily usage of the two words occupies 29.51\% and 42.79\% of their referring word ``chinese'' in the pre-pandemic period.
However, this proportion climbs to 41.77\% and 52.75\% in the pandemic period, indicating users are inclined to use slur words to refer to Chinese people.
We see a different slur distribution on Reddit (see Figure~\ref{figure: slur_reddit}).
Since Reddit establishes hate-speech ban policy from Jun, 2020~\cite{Reddit_ban_hate}, the frequency of Sinophobic slurs does not rise up as significantly as /pol/, but we still observe slurs sneak on Reddit.
For instance, ``chicoms'' increases 39.73\% in the pandemic period.
These findings highlight the wildly usage of Sinophobic slurs on both 4chan's /pol/ and Reddit, especially during the pandemic period.

\begin{figure*}[t]
    \centering
	\begin{subfigure}{0.32\linewidth}
    \centering
		\includegraphics[width=\linewidth]{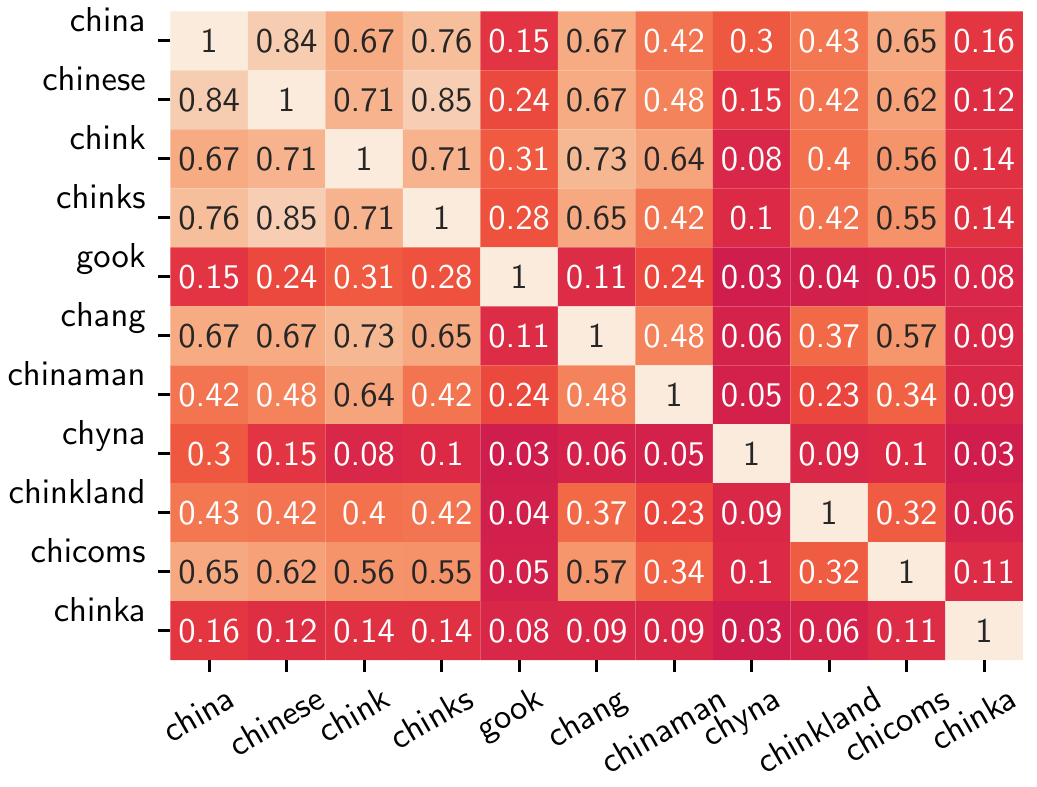}
	\caption{Pre-pandemic period on 4chan's /pol/}
	\label{figure: 4chan_perspective_pre}
	\end{subfigure}\hfill
    \begin{subfigure}{0.32\linewidth}
        \centering
		\includegraphics[width=\linewidth]{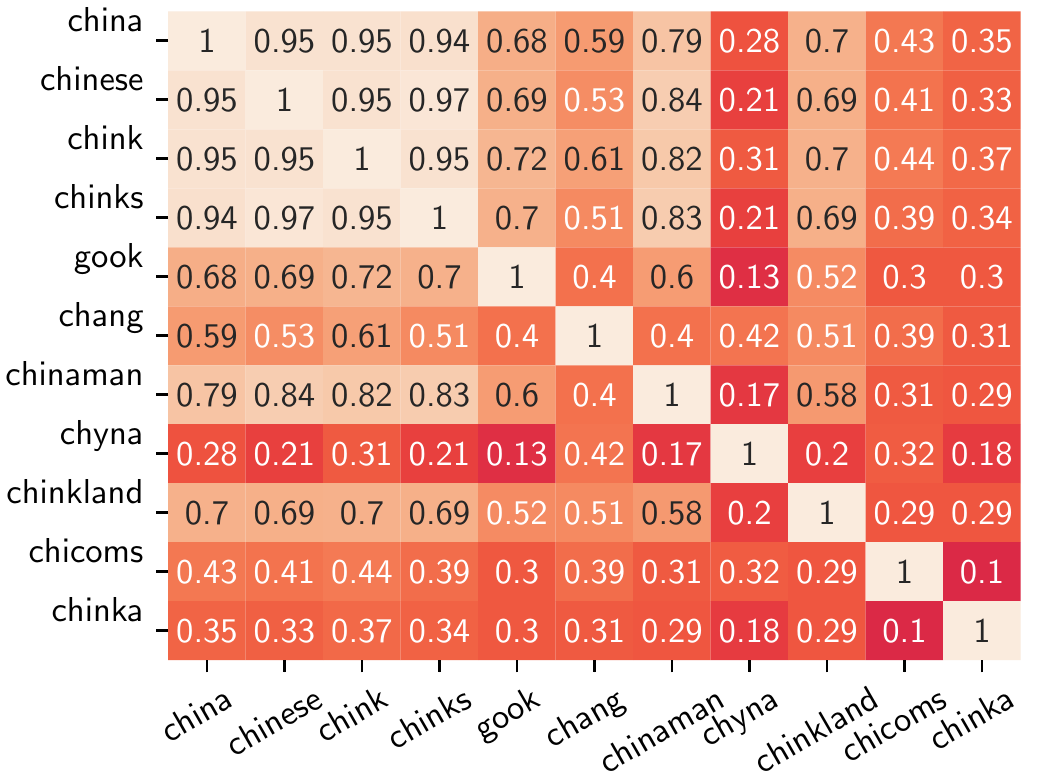}
		\caption{Pandemic period on 4chan's /pol/}
    \end{subfigure}\hfill
    \begin{subfigure}{0.32\linewidth}
        \centering
		\includegraphics[width=\linewidth]{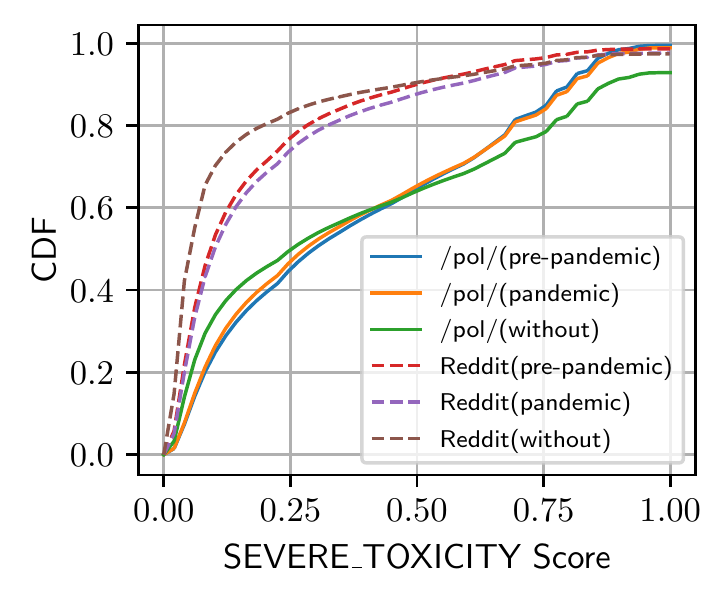}
		\caption{SEVERE\_TOXICITY}
		\label{figure: severe_toxicity}
    \end{subfigure}\hfill

	\begin{subfigure}{0.32\linewidth}
    \centering
		\includegraphics[width=\linewidth]{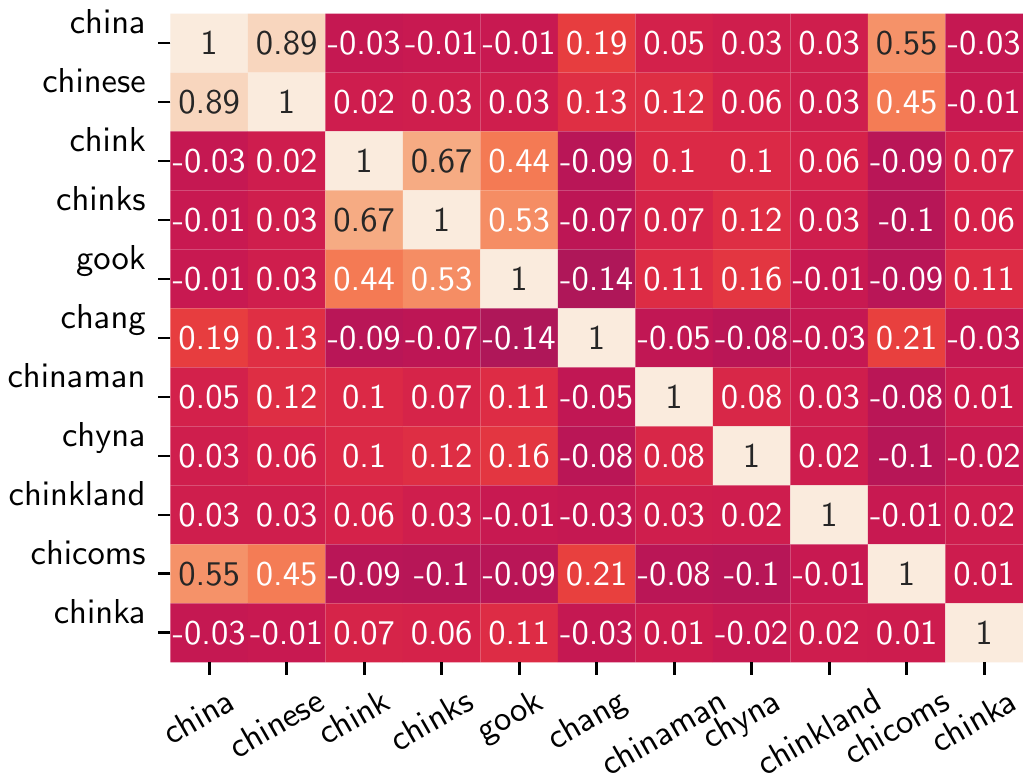}
		\caption{Pre-pandemic period on Reddit}
	\end{subfigure}\hfill
    \begin{subfigure}{0.32\linewidth}
        \centering
		\includegraphics[width=\linewidth]{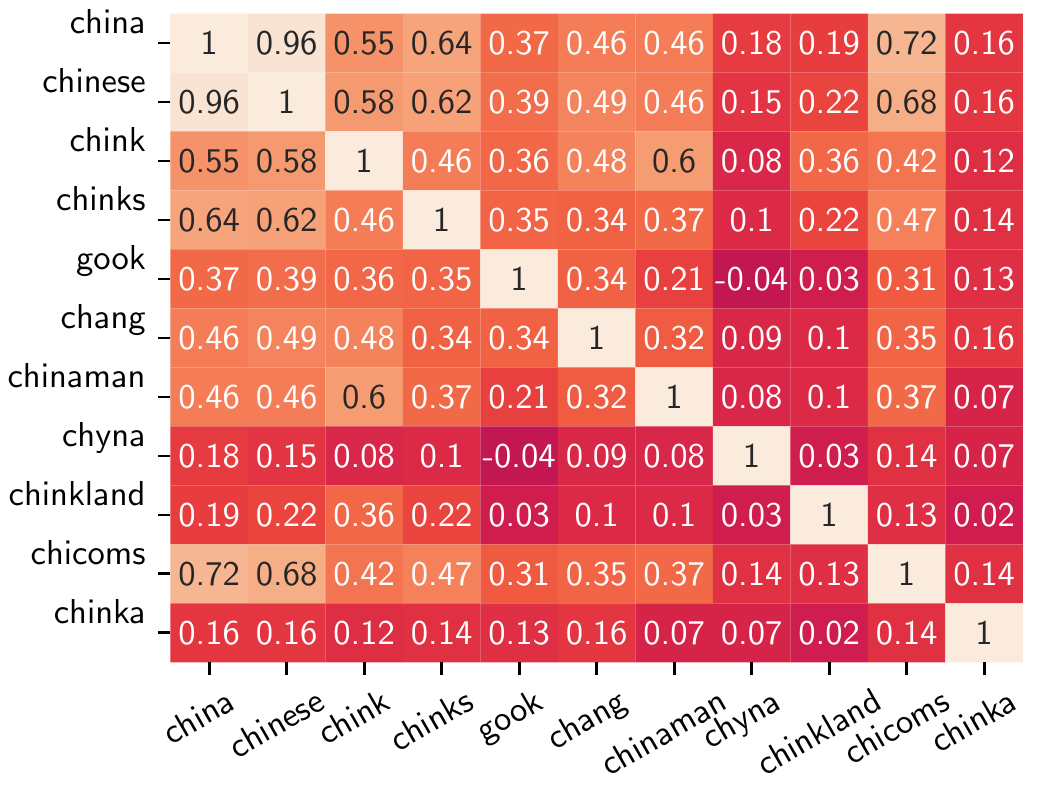}
		\caption{Pandemic period on Reddit}
    \end{subfigure}\hfill
    \begin{subfigure}{0.32\linewidth}
        \centering
		\includegraphics[width=\linewidth]{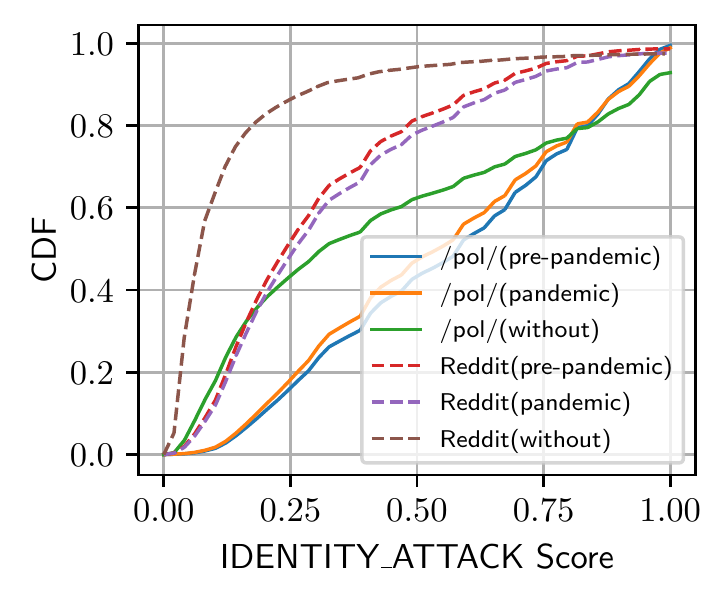}
		\caption{IDENTITY\_ATTACK}
		\label{figure: indentity_attack}
    \end{subfigure}\hfill
\caption{In this figure, (a) and (b) ((d) and (e)) show the correlation coefficient between the percentage trend of different Sinophobic slurs in the pre-pandemic and pandemic period on 4chan's /pol/ (Reddit). 
(c) ((f)) shows the CDFs of SEVERE\_TOXICITY (IDENTITY\_ATTACK). 
Here (pre-pandemic)/(pandemic) means the posts that contain ``china'' or ``chinese'' in the pre-pandemic/pandemic period and (without) means the posts that do not contain ``china'' or ``chinese'' in the whole period.}
\label{figure: cc_slurs_reddit}
\end{figure*}

\subsection{Correlation}

To verify the tendency relationship between the term ``china'', ``chinese'', and slur words in a quantitative way, we measure the correlation coefficient of them in different periods.
Concretely, for each Web community, we split the percentage values of daily posts into pre-pandemic and pandemic periods. 
We then calculate the Pearson correlation among the pre-pandemic percentage values and pandemic percentage values of the eleven terms, i.e., ``china,'' ``chinese,''as well as nine Sinophobic slurs (see Figure~\ref{figure: cc_slurs_reddit}).

We find the correlation between ``china'' and ``chinese'' is close on 4chan's /pol/ (0.84) and Reddit (0.89) in the pre-pandemic period, and after COVID-19 outbreaks, it even mounts to 0.95 and 0.96 on /pol/ and Reddit, respectively,  which sheds light on the increasing homogenization among the country and people.
In addition, official words, e.g. ``china, ``chinese'', also obtain a close-knit relationship to most of ethnic slurs in the pandemic period. 
Take ``chinaman'' and its referring term ``chinese'' as an example, in the pre-pandemic period, the correlation of ``chinaman'' is 0.48 on 4chan's /pol/ and 0.12 on Reddit.
However, in the pandemic period, it surges to 0.84 on 4chan's /pol/ and 0.46 on Reddit, which indicates that the discussions around China and Chinese people substantially changed during the pandemic period.

\subsection{Perspective Analysis}
\label{section:perspective_analysis}

We then use Google's Perspective API~\cite{GooglePerspectiveAPI} to identify two kinds of perspectives towards the posts, i.e., SEVERE\_TOXICITY and IDENTITY\_ATTACK, in which SEVERE\_TOXICITY perspective measures the degree of hate, rudeness, and disrespect of comments and IDENTITY\_ATTACK evaluates how negative or hateful that the comments are targeting someone because of their identity.
Concretely, for each Web community, we select the posts that contain ``china'' or ``chinese'' in the pre-pandemic and pandemic period.
Note that we also extract 1\% of all posts without the term ``china'' or ``chinese'' as a baseline.
The CDF of different scores are summarized in Figure~\ref{figure: severe_toxicity} and Figure~\ref{figure: indentity_attack} and we list our findings as follows.

First, we observe that /pol/ posts are more toxic than Reddit in general, which is expected as /pol/ users are anonymous and considered more notorious~\cite{PZCSB20}, and the platform is less moderated.
For instance, the percentage of posts with SEVERE\_TOXICITY score greater than 0.5 is more than 34.26\% on 4chan's /pol/, while only 10.79\% on Reddit.
Second, compared to the posts without mentioning ``china'' and ``chinese'', the posts containing them have a higher score across all Perspective dimensions.
Take /pol/ as an example, the percentage of posts with IDENTITY\_ATTACK score $\geq$ 0.5 is more than 53.91\% for the posts containing ``china'' or ``chinese,'' while only 37.16\% for the posts without them. 
Third, compared to the pre-pandemic period, the pandemic period has a higher toxicity level.
For example, in the pre-pandemic period, the percentage of posts with SEVERE\_TOXICITY $\geq$ 0.8 is 3.15\% on Reddit, and it climbs to 4.65\% in the pandemic period.

\mypara{Takeaways}
We find that events that are not directly related to China may also evoke discussion to China and Chinese people, e.g., the United States withdraw from the Paris Agreement~\cite{USWITHDRAWPARIS} and Inauguration of Joe Bide~\cite{BIDENINAUGURATION}.
And the discussions around China and Chinese people are prolonged in the pandemic period.
For instance, in the pre-pandemic period, the average interest time of ``china'' (``chinese'') remains 2.76 (2.83) days while it increases to 8.00 (9.28) days in the pandemic period.
Analysis of Sinophobic slurs reveals that they are widely used on both fringe and mainstream Web community, especially during the pandemic period.
Surprisingly, even with the hate-speech ban policy~\cite{Reddit_ban_hate}, slurs have still been observed sneaked on Reddit, which calls for the mainstream community to take more actions and responsibility.
We also find the usage patterns of Sinophobic slurs become more similar on both Web communities in the pandemic period, implying that people's linguistic habits referring to the Chinese are tilting towards slurs. 
Perspective analysis towards Sinophobic slurs shows that Sinophobia is more severe in posts from 4chan's /pol/, posts that contain ``china'' or ``chinese'', and posts in the pandemic period.

\section{Content Analysis}

In this section, we aim to study the evolution of Sinophobia from the content of the posts in our dataset.
Specifically, we measure how drastically the semantic meaning of ``china'' and ``chinese'' has changed and understand the shift in a diachronic visualization way.
Next, we provide a detailed analysis on a number of Sinophobic topics discussed on 4chan's /pol/ and Reddit over six years.

\subsection{Semantic Evolution}

To have better speculation on word meaning shifting of ``china'' and ``chinese'', for each Web community, we train multiple word2vec models on corpus of each month with the same setting we used to train the whole model (see the section ``Temporal Analysis'').
Note that we discard June, 2016 as it only contains two days' posts on /pol/, which is not meaningful for comparison.
In this way, for each Web community, we have 57 word2vec models corresponding to each month from July, 2016 to March, 2021.
We treat the pre-trained GoogleNews model~\cite{MCCD13} released by Google as our baseline model and align monthly models to it to ensure that the vectors are projected to the same coordinate axes~\cite{HLJ16}.

\begin{figure*}[htbp]
  \centering
  	\begin{subfigure}{.16\linewidth}
		\centering
		\includegraphics[width= \linewidth]{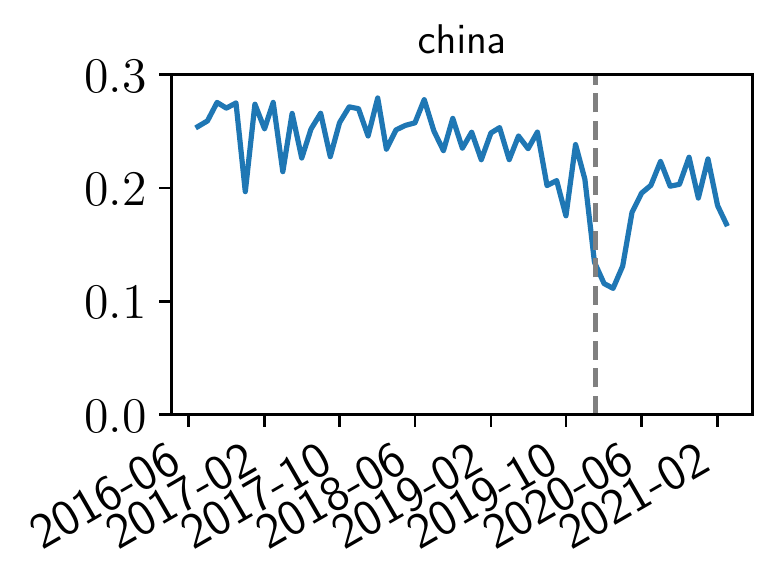}
	\end{subfigure}\hfill
	\begin{subfigure}{.16\linewidth}
    	\centering 
    	\includegraphics[width=\textwidth]{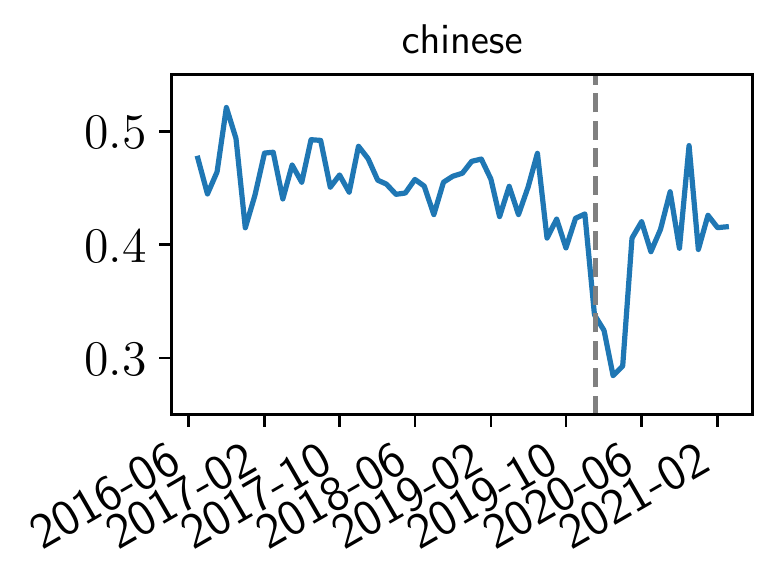}
    \end{subfigure}\hfill
	\begin{subfigure}{.16\linewidth}
    	\centering 
    	\includegraphics[width=\textwidth]{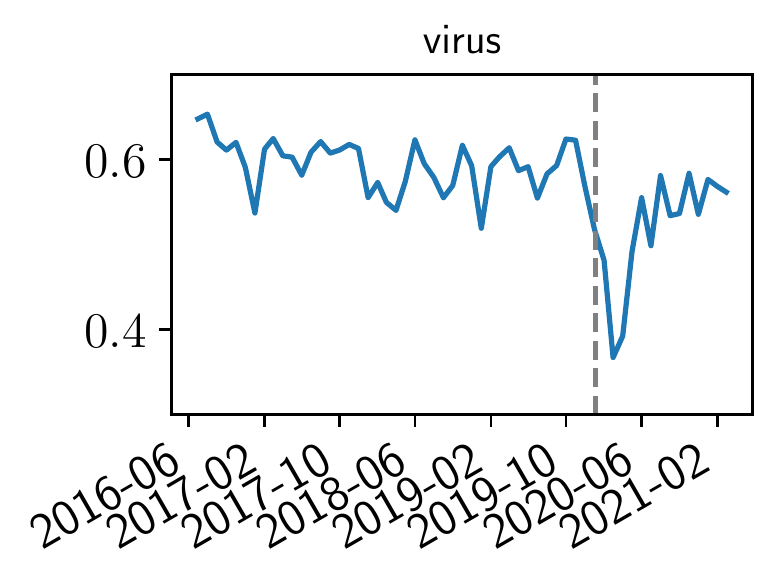}
    \end{subfigure}\hfill
  	\begin{subfigure}{.16\linewidth}
		\centering
		\includegraphics[width= \linewidth]{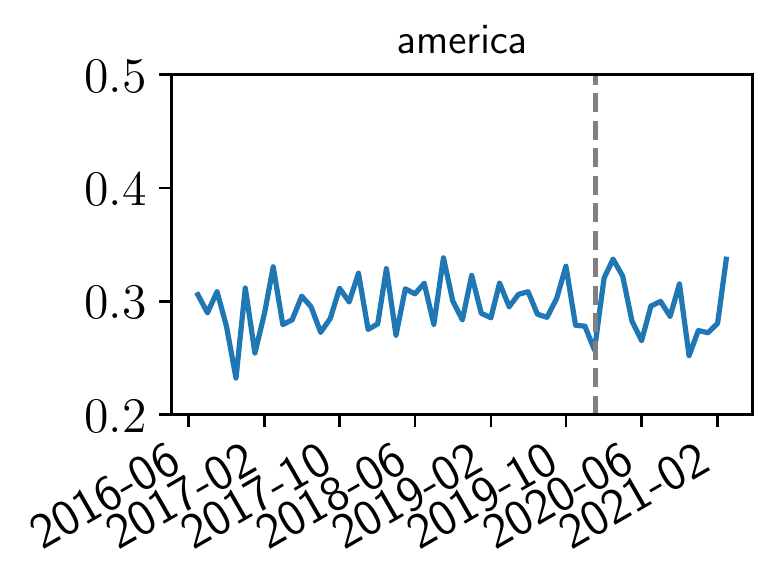}
	\end{subfigure}\hfill
	\begin{subfigure}{.16\linewidth}
    	\centering 
    	\includegraphics[width=\textwidth]{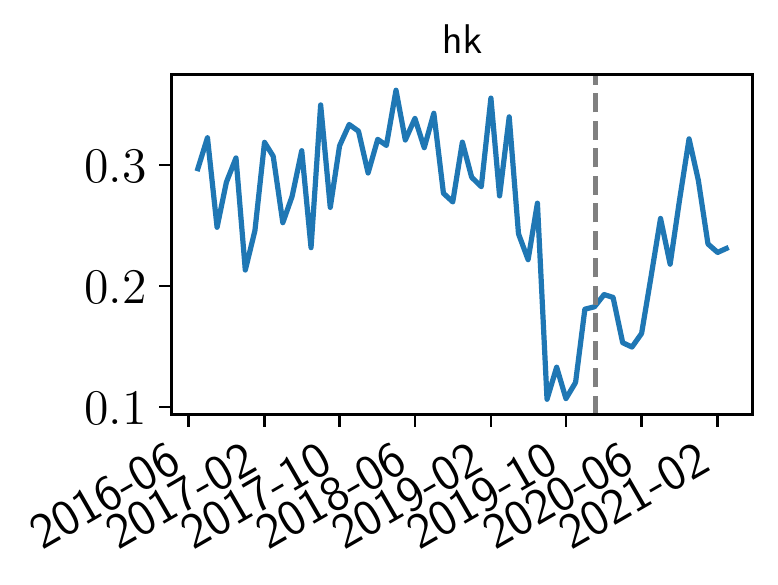}
    \end{subfigure}\hfill
	\begin{subfigure}{.16\linewidth}
    	\centering 
    	\includegraphics[width=\textwidth]{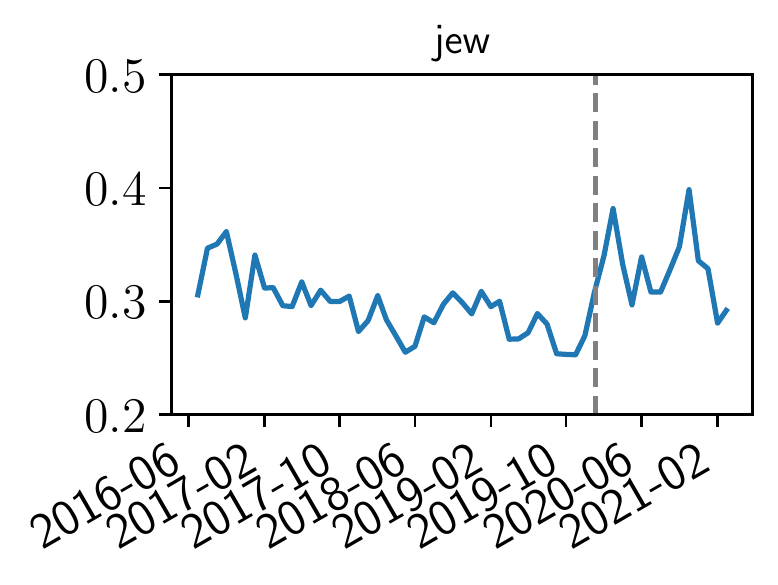}
    \end{subfigure}\hfill

\caption{Word meaning changes on 4chan's /pol/. Y-axis is cosine similarity between vectors of month models and the pre-trained GoogleNews model. The grey line corresponds to the split month (Jan, 2020).}
\label{figure:word2vec_meaning_change_pol}
\end{figure*}

\begin{figure*}[htbp]
  \centering
  	\begin{subfigure}{.16\linewidth}
		\centering
		\includegraphics[width= \linewidth]{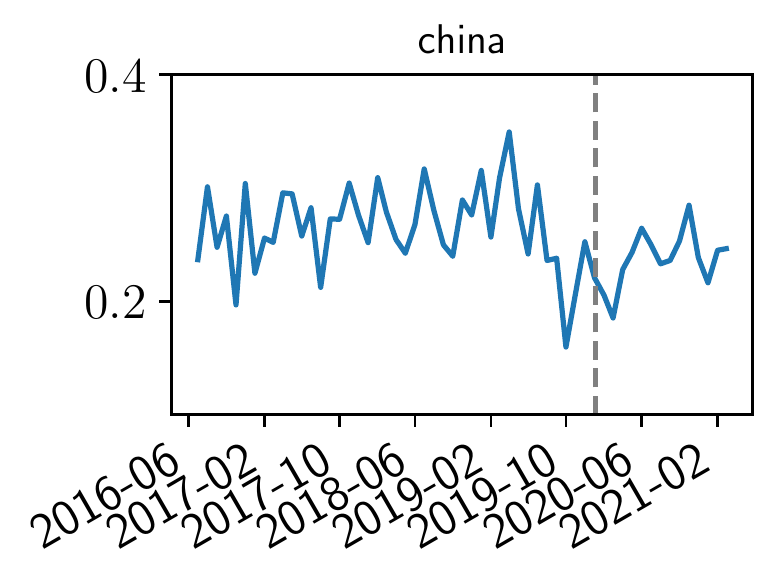}
	\end{subfigure}\hfill
	\begin{subfigure}{.16\linewidth}
    	\centering 
    	\includegraphics[width=\textwidth]{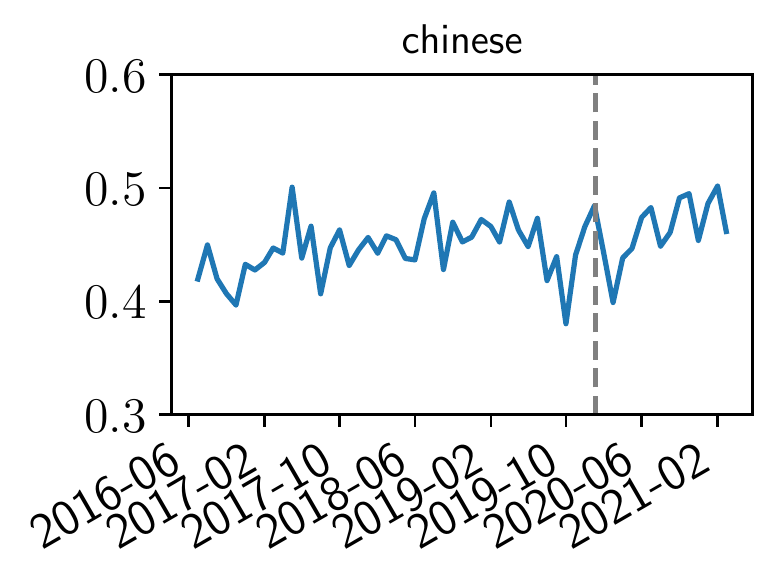}
    \end{subfigure}\hfill
	\begin{subfigure}{.16\linewidth}
    	\centering 
    	\includegraphics[width=\textwidth]{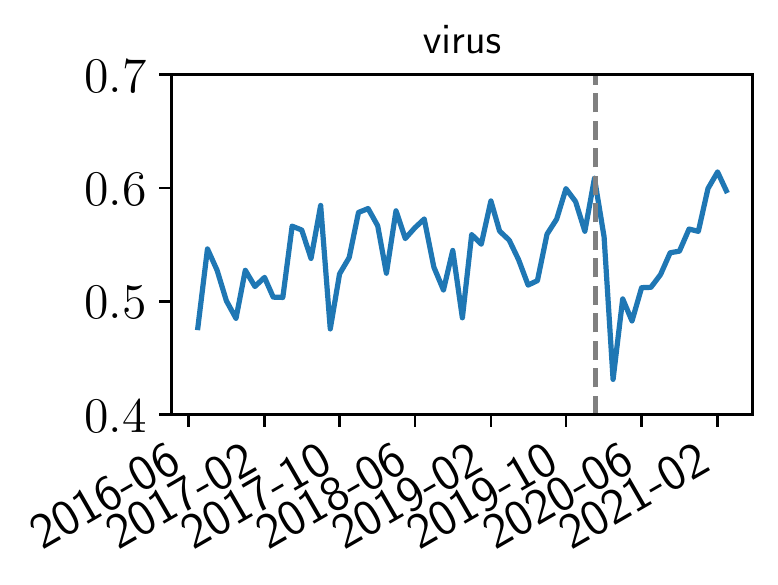}
    \end{subfigure}\hfill
  	\begin{subfigure}{.16\linewidth}
		\centering
		\includegraphics[width= \linewidth]{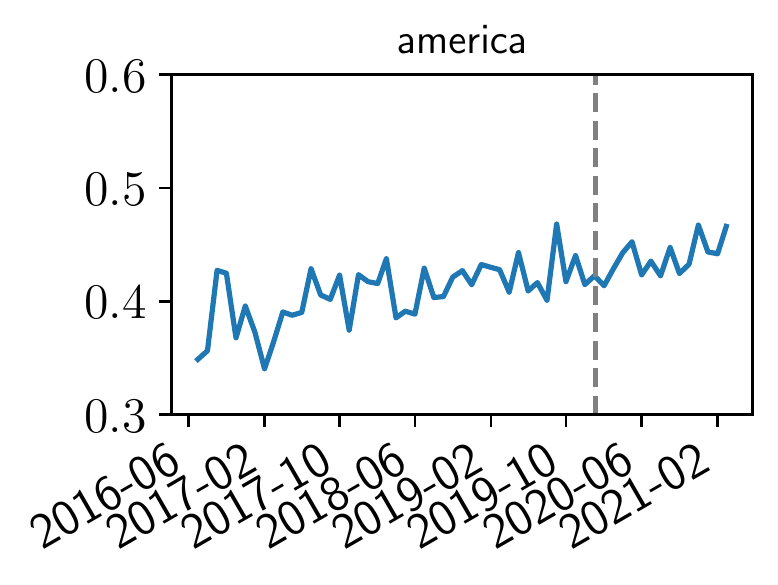}
	\end{subfigure}\hfill
	\begin{subfigure}{.16\linewidth}
    	\centering 
    	\includegraphics[width=\textwidth]{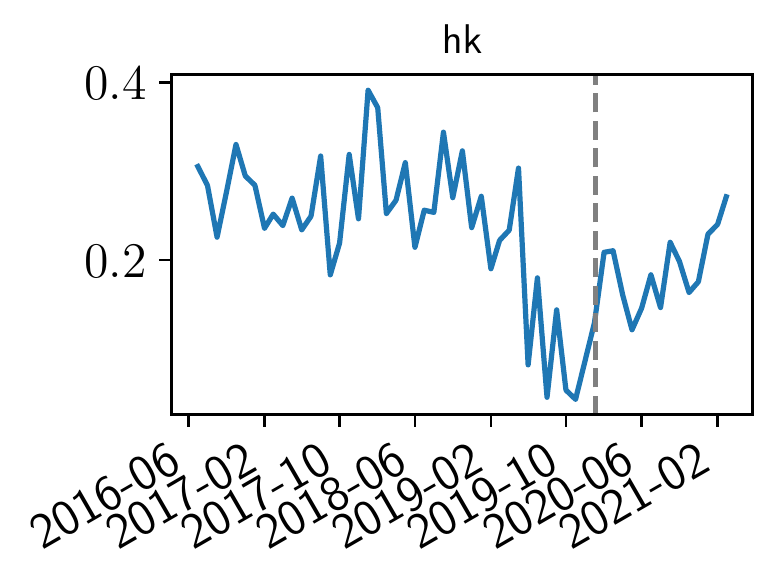}
    \end{subfigure}\hfill
	\begin{subfigure}{.16\linewidth}
    	\centering 
    	\includegraphics[width=\textwidth]{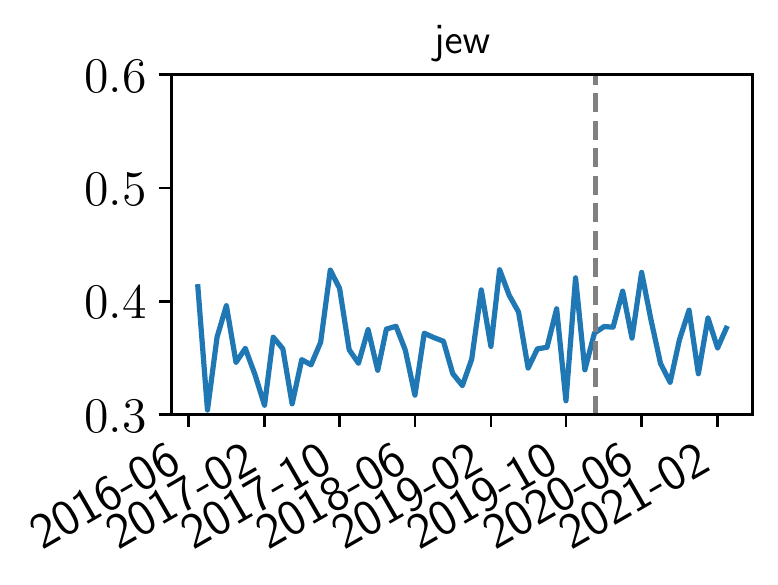}
    \end{subfigure}\hfill
\caption{Word meaning changes on Reddit. Y-axis is cosine similarity between vectors of month models and the pre-trained GoogleNews model.  The grey line corresponds the split month (Jan, 2020).}
\label{figure:word2vec_meaning_change_reddit}
\end{figure*}

Figure~\ref{figure:word2vec_meaning_change_pol} and Figure~\ref{figure:word2vec_meaning_change_reddit} display the diachronic word embeddings of ``china,'' ``chinese,'' ``virus,'' ``hk,'' ``america,'' and ``jew'' on 4chan's /pol/ and Reddit, respectively.
We choose ``virus'' and ``hk'' to be the baseline of meaning shift towards events, ``america'' as a comparison of ``china'', and ``jew'' to compare with ``chinese''.
The Y-axis is the cosine similarity between the vector of monthly models (aligned) and the baseline model, which is a conventional evaluation metric used to measure word embedding differences~\cite{ZFBB20, TSLBSZZ21}.
First, we observe the most significant semantic changes of these words happen in different periods.
For ``china'', ``chinese'', and ``virus'', it is during COVID-19 outbreaks (Jan, 2020); for ``america'', it happens when Donald Trump wins the U.S. election (Nov, 2016); for ``hk'', it is during Hong Kong protests (Oct, 2019); and for ``jew'', it is when Synagogue attack happened at High Holy Day (Oct, 2019).
These findings hold for both 4chan's /pol/ and Reddit, which indicate that the semantic changes correspond to political events and are a cross-platform phenomenon.
We also measure the average cosine similarity for different periods.
Specifically, in the pre-pandemic period, the average cosine similarity of ``china'' (``chinese'') is 0.25 (0.46) on 4chan's /pol/ with variance of 0.02 (0.03). 
However, in the pandemic period, this value shifts to 0.18 (0.39) with variance of 0.04 (0.06), which indicates more significant semantic changes after COVID-19. 
To compare, the average cosine similarity of ``hk'', ``america'', and ``jew'' is 0.28, 0.30, 0.29 with variance of 0.07, 0.02, and 0.03 in the pre-pandemic period and 0.22, 0.29, and 0.33 with variance of 0.05, 0.03, and 0.03 in the pandemic period, which is more stable than ``china'', ``chinese'', and ``virus''.

\begin{figure*}[htbp]
    \centering
    \includegraphics[width=0.95\linewidth]{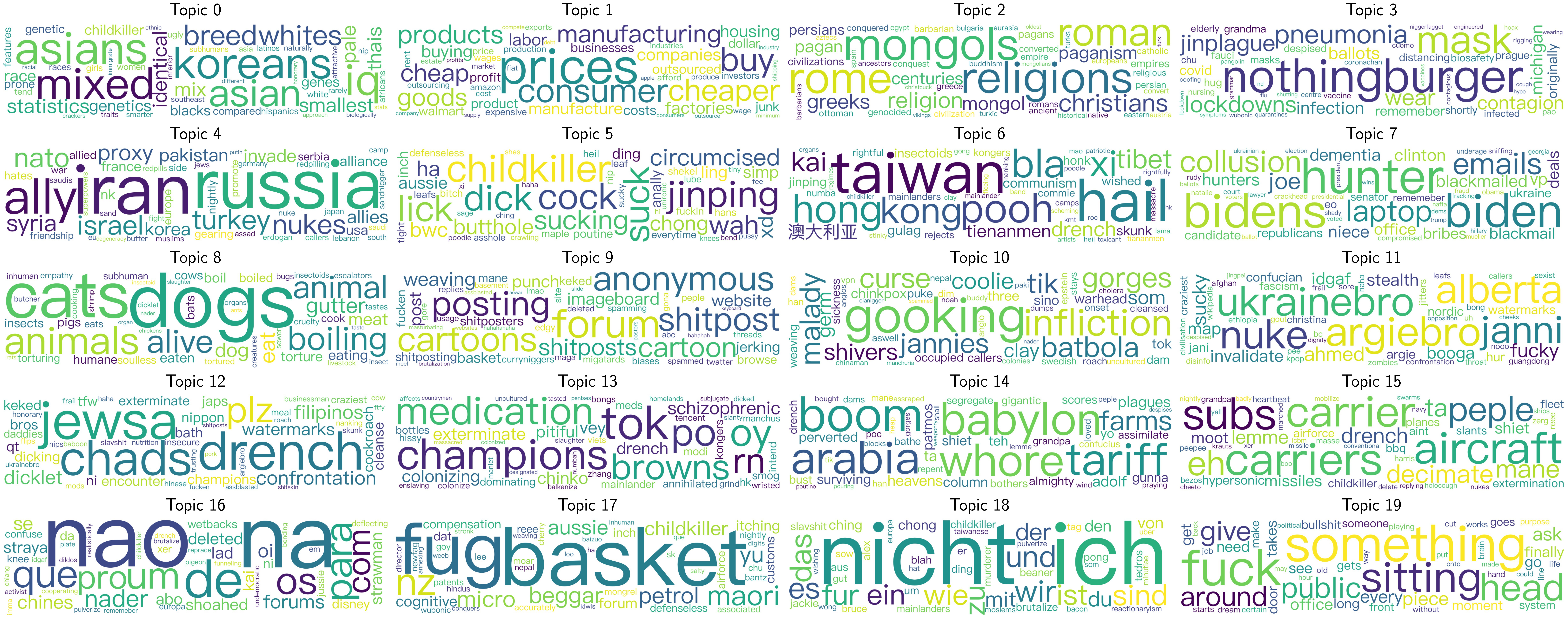}
\caption{ Sinophobic-related topics on 4chan's /pol/.}
\label{figure: topic_word_cloud_pol}
\end{figure*}

\begin{figure*}[htbp]
		\centering
		\includegraphics[width=0.95\linewidth]{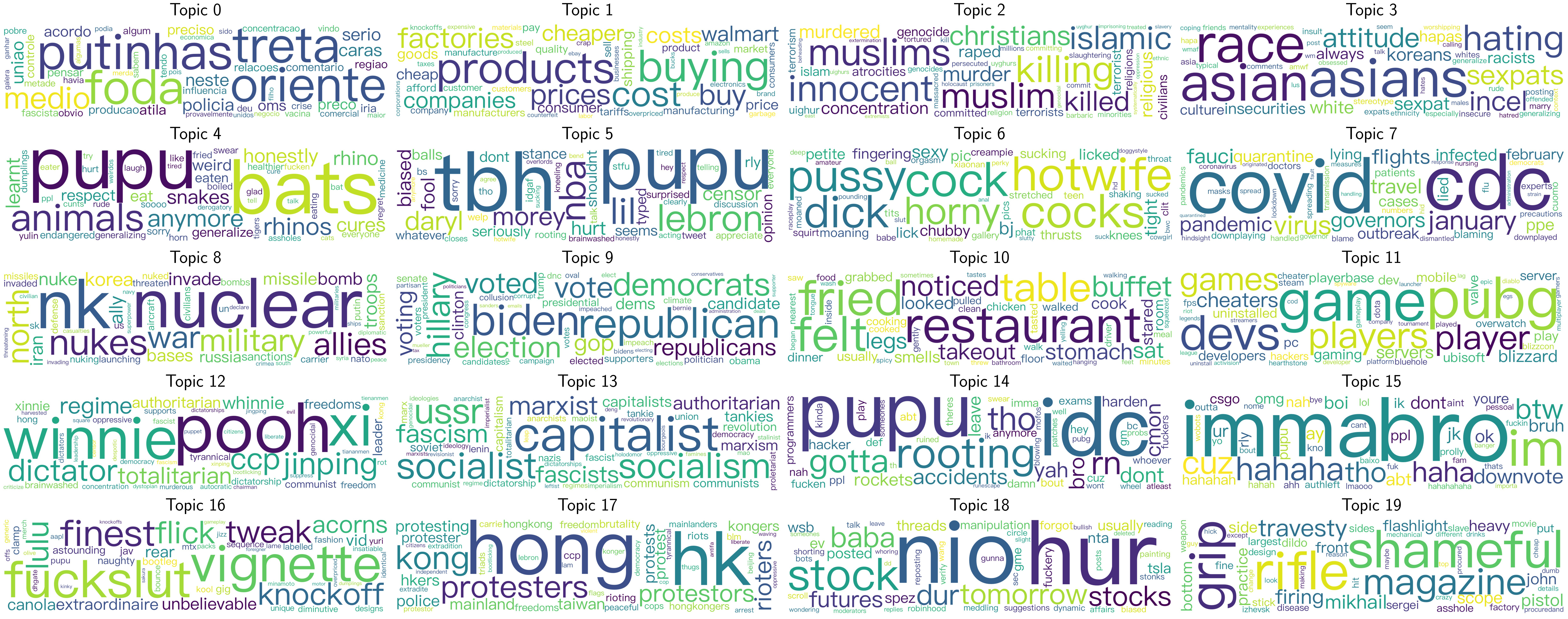}
\caption{ Sinophobic-related topics on Reddit.}
\label{figure: topic_word_cloud_reddit}
\end{figure*}

\begin{figure}[t]
\centering
  \begin{subfigure}{0.45\linewidth}
    \centering
		\includegraphics[width=\linewidth]{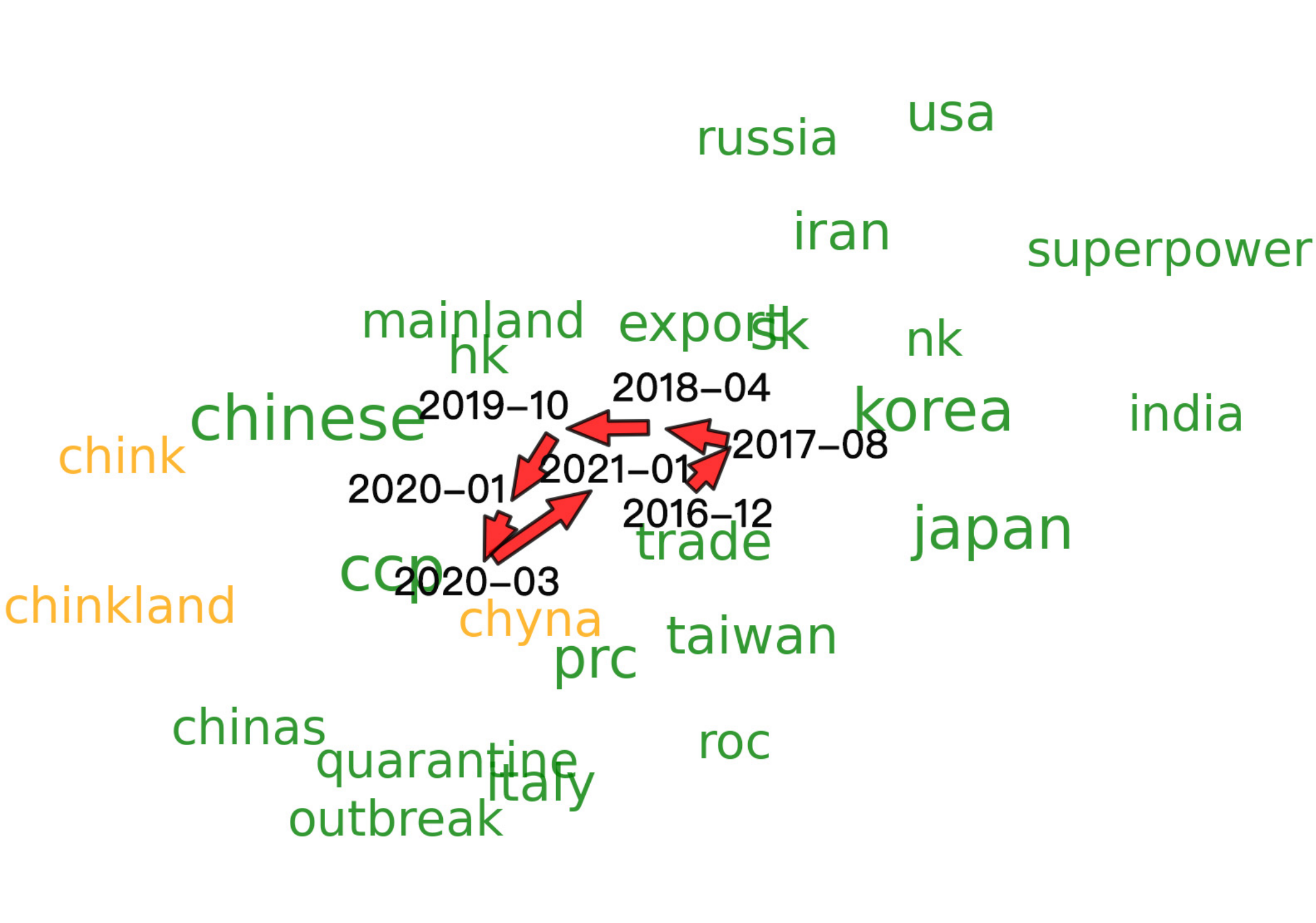}
		\caption{``china'' on 4chan's /pol/}
		\label{figure:histword_china_pol}
    \end{subfigure}\hfill
      \begin{subfigure}{0.45\linewidth}
		\centering
		\includegraphics[width=\linewidth]{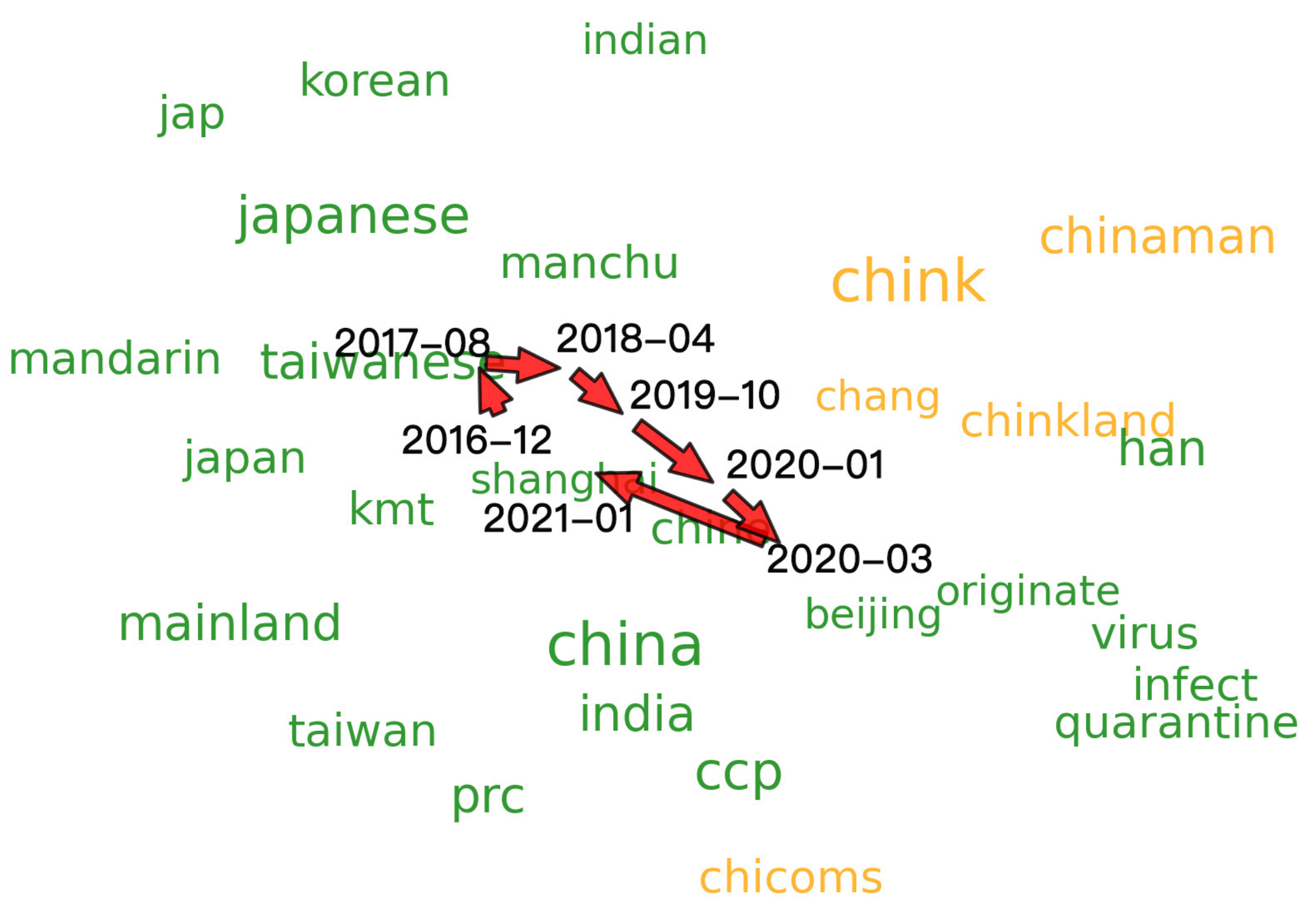}
		\caption{``chinese'' on 4chan's /pol/}
		\label{figure:histword_chinese_pol}
\end{subfigure}\hfill
  \begin{subfigure}{0.45\linewidth}
		\centering
		\includegraphics[width=\linewidth]{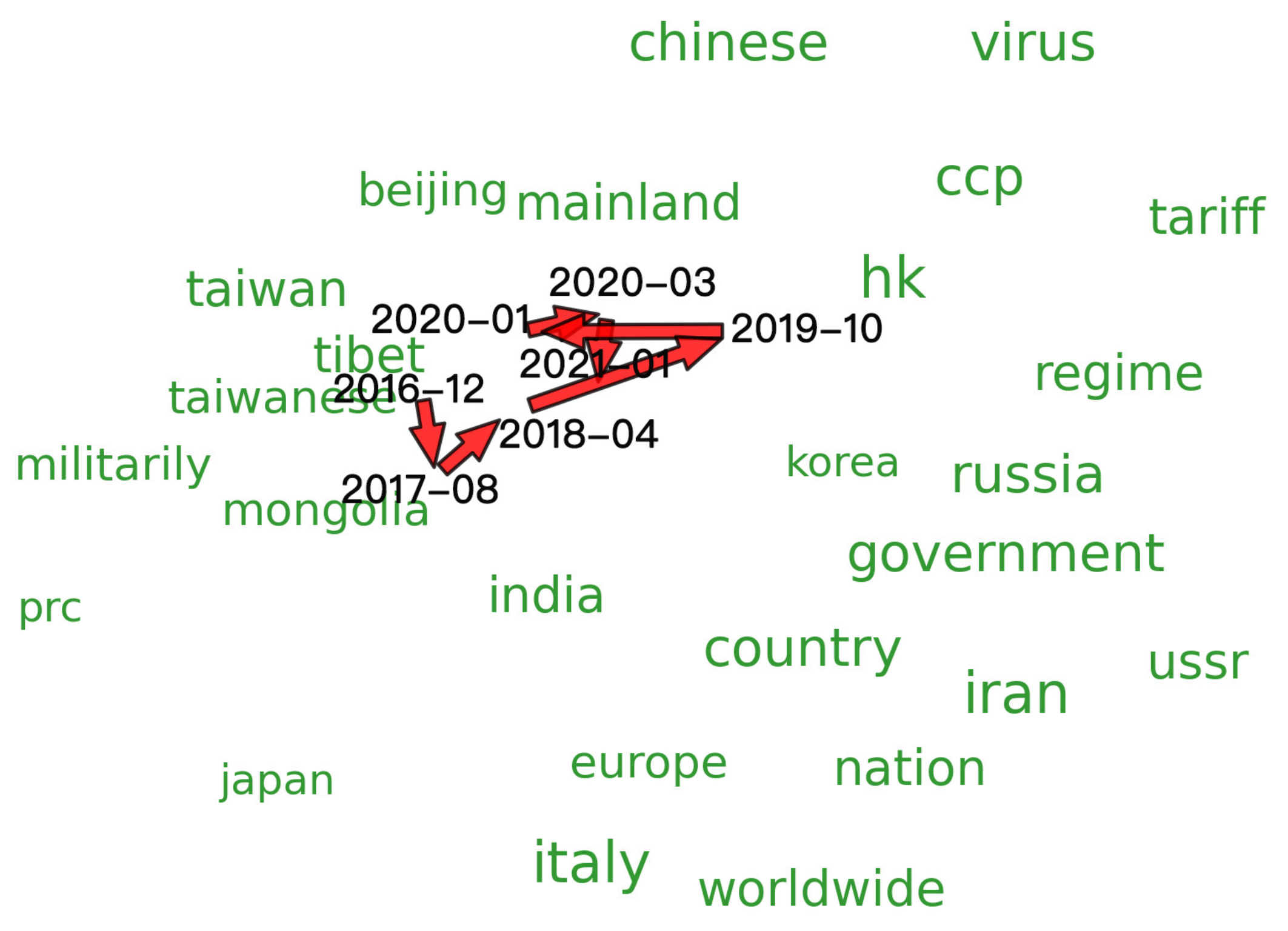}
		\caption{``china'' on Reddit}
		\label{figure:histword_china_reddit}
	\end{subfigure}\hfill
  	\begin{subfigure}{0.45\linewidth}
		\centering
		\includegraphics[width=\linewidth]{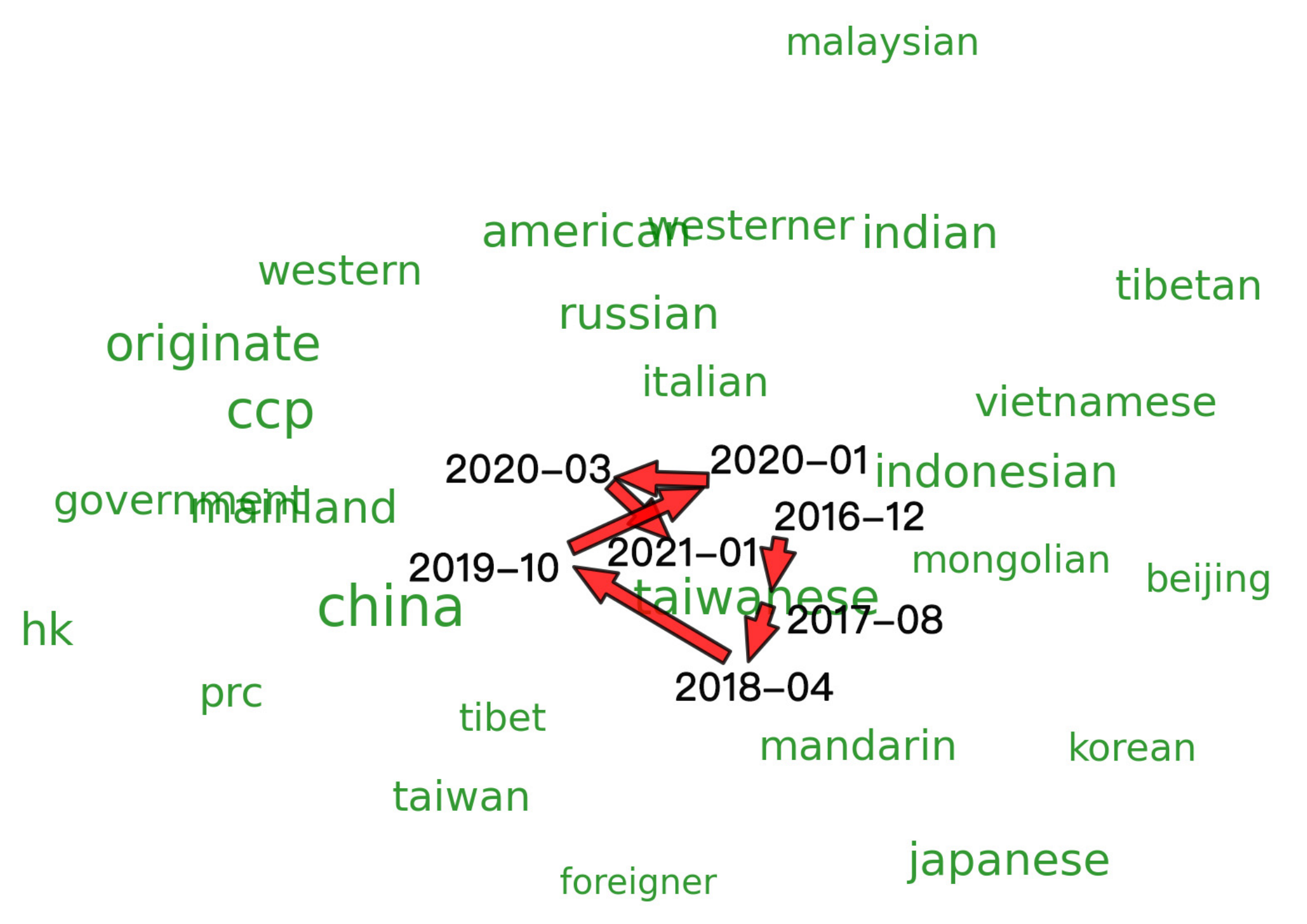}
		\caption{``chinese'' on Reddit}
		\label{figure:histword_chinese_reddit}
	\end{subfigure}\hfill
\caption{Diachronic word embeddings of words in 7 months on 4chan's /pol/ and Reddit. Sinophobic slurs are annotated in orange color.}
\label{figure: histwords}
\end{figure}

\subsection{Analysis of Diachronic Word Embeddings}

To further understand the meaning behind such dramatic changes, we visualize the diachronic word embeddings of words that are similar to the term ``china'' and ``chinese'', following the methodology proposed by~\cite{HLJ16}.
In a nutshell, we start by selecting the month including the split month (2020-01) and six other months with the highest frequent mentions to ``china'' and ``chinese'' for each year, i.e., 2016-12, 2017-08, 2018-04, 2019-10, 2020-03, and 2021-01 (see  Figure~\ref{figure:china_chinese_occ_per}).

Next, we train the word2vec model for each month and align these word2vec models to the first one (2016-12).
All embeddings are projected into two dimensions via t-SNE~\cite{MH08}.
For each model, we select the top 10 most similar words (denoted as reference words) of the keyword ``china'' or ``chinese'' and the results are depicted in Figure~\ref{figure: histwords}.

By inspecting the positions of words in Figure~\ref{figure: histwords}, we can measure the corresponding meaning shifting in each month.
For instance, for ``china'' on 4chan's /pol/ (see Figure~\ref{figure:histword_china_pol}), the meaning is close to ``taiwan'' and ``asia'' in November 2016.
Then, it moves to ``nk'', ``sk'', and ``korea'' in August 2017, ``prc'' in April 2018, ``hk'' and ``mainland'' in October 2019.
Until 2019, all these shifts are still related to geographic terms and political events, e.g., Trump–Tsai Call~\cite{TrumpTsaicall}, North Korean Nuclear Test~\cite{NKNUCLEARTEST}, Hong Kong Protests~\cite{HKPROTESTS}.
However, when it comes to January 2020, the meaning of ``china'' not only shift to pandemic words, e.g., ``quarantine'' and ``outbreak'', but also towards slur words like ``chink' and ``chinkland''.
Then, it moves to ``chyna'' in March 2020, and finally takes a step back in January 2021, which shows that the meaning of ``china'' is shifting from the notation of country to the Sinophobic slur words.
We have a similar observation for ``chinese'' on /pol/.
For Reddit, we find that they also shift from region-related words, e.g., ``taiwan'' and ``tibet'' to pandemic words, e.g., ``tariff'' and ``virus.'' 
However, we do not observe Sinophobic slur words on it.

\subsection{Topic Model}

\begin{table*}[!t]
\centering
\small{
\begin{tabular}{rlrrr}
\toprule
\multicolumn{1}{l}{\textbf{Topic}} &
  \textbf{Topic Words} &
  \multicolumn{1}{l}{\textbf{Pre(\%)}} &
  \multicolumn{1}{l}{\textbf{Post (\%)}} &
  \multicolumn{1}{l}{ \textbf{Ratio}} \\
  \midrule
0  & asians, koreans, mixed, iq, asian               & \textbf{9.21} & 4.46 & 0.48 \\
1  & prices, consumer, buy, cheaper, products        & 7.68          & 5.28 & 0.69 \\
2  & mongols, religions, rome, roman, christians     & 8.4           & 3.91 & 0.47 \\
3  & nothingburger, mask, pneumonia, jinplague, lockdowns & 0.97 & \textbf{10.67} & \textbf{11.00} \\
4  & iran, russia, ally, turkey, nukes              & 6.58 & 5.1 & 0.78 \\
5  & childkiller, jinping, dick, cock, lick         & 4.76          & 5.94 & 1.25 \\
6  & hail, taiwan, pooh, hong, bla                   & 5.31          & 5.41 & 1.02 \\
7  & hunter, bidens, biden, collusion, emails        & 4.04          & 6.44 & 1.59 \\
8  & dogs, cats, animals, boiling, alive             & 4.93          & 5    & 1.01 \\
9  & forum, anonymous, posting, cartoons, shitpost   & 4.8           & 4.93 & 1.03 \\
10 & gooking, infliction, gorges, tik, curse         & 4.65          & 4.89 & 1.05 \\
11 & ukrainebro, argiebro, nuke, alberta, janni      & 4.33          & 5.07 & 1.17 \\
12 & jewsa, drench, chads, plz, confrontation        & 4.65          & 4.73 & 1.02 \\
13 & tok, champions, medication, oy, po              & 5.03          & 3.97 & 0.79 \\
14 & boom, babylon, arabia, tariff, whore            & 4.51          & 4.16 & 0.92 \\
15 & carriers, aircraft, subs, carrier, peple        & 4.52          & 4.05 & 0.90 \\
16 & na, nao, de, que, para                          & 4.11          & 4.38 & 1.07 \\
17 & basket, fug, nz, maori, beggar                  & 4.22          & 3.86 & 0.91 \\
18 & nicht, ich, und, das, wir                       & 4.27          & 3.78 & 0.89 \\
19 & something, fuck, sitting, head, public          & 3.02          & 3.97 & 1.31 \\
\bottomrule
\end{tabular}
}
\caption{Percentage of each topic in pre-pandemic and pandemic period on 4chan's /pol/. Topic words are the top 5 weighted words in the topic. Pre represents the pre-pandemic period. Post refers to the pandemic  period. Ratio is the ratio of Post to Pre.}
\label{table:topic_per_4chan}
\end{table*}

\begin{table*}[!t]
\centering
\small
{
\begin{tabular}{rlrrr}
\toprule
\multicolumn{1}{l}{\textbf{Topic}} &
  \textbf{Topic Words} &
  \multicolumn{1}{l}{\textbf{Pre(\%)}} &
  \multicolumn{1}{l}{\textbf{Post(\%)}} &
  \multicolumn{1}{l}{\textbf{Ratio}} \\
  \midrule
0  & putinhas, treta, oriente, foda, medio          & \textbf{12.41} & \textbf{14.85} & 1.20 \\
1  & products, buying, cost, buy, prices             & 6.82 & 4.77 & 0.70          \\
2  & muslims, killing, innocent, muslim, islamic     & 5.84 & 5.55 & 0.95          \\
3  & asians, race, asian, hating, sexpats            & 6.59 & 4.6  & 0.70          \\
4  & bats, pupu, animals, anymore, rhinos            & 5.11 & 5.85 & 1.14          \\
5  & tbh, pupu, lebron, lil, nba                     & 5.09 & 5.57 & 1.09          \\
6  & cock, hotwife, cocks, pussy, dick               & 5.11 & 4.86 & 0.95          \\
7  & cdc, covid, january, virus, governors           & 1.53 & 8.14 & \textbf{5.32} \\
8  & nuclear, nk, nukes, military, allies            & 5.32 & 4.01 & 0.75          \\
9  & biden, republican, democrats, hillary, election & 4.52 & 4.56 & 1.01          \\
10 & restaurant, fried, felt, table, noticed         & 5.55 & 3.15 & 0.57          \\
11 & game, pubg, devs, players, player               & 5.61 & 3.01 & 0.54          \\
12 & pooh, winnie, xi, jinping, dictator             & 3.12 & 5.33 & 1.71          \\
13 & capitalist, socialist, ussr, socialism, marxist & 3.89 & 4.36 & 1.12          \\
14 & idc, pupu, rooting, gotta, rn                   & 4.65 & 3.41 & 0.73          \\
15 & bro, imma, im, hahaha, cuz                      & 3.3  & 4.58 & 1.39          \\
16 & vignette, fuckslut, finest, tweak, knockoff     & 4.2  & 3.53 & 0.84          \\
17 & hong, hk, kong, protesters, protestors          & 4.25 & 3.26 & 0.77          \\
18 & hur, nio, stock, tomorrow, baba                 & 3.89 & 3.59 & 0.92          \\
19 & rifle, shameful, grip, magazine, travesty       & 3.19 & 3.02 & 0.95         \\
\bottomrule
\end{tabular}
}
\caption{Percentage of each topic in pre-pandemic and pandemic period on Reddit. Topic words are the top 5 weighted words in the topic. Pre represents the pre-pandemic period. Post refers to the pandemic  period. Ratio is the ratio of Post to Pre.}
\label{table:topic_per_reddit}
\end{table*}

We then dive into the detailed toxic topics that are related to China and Chinese in 4chan's /pol/ and Reddit.

Concretely, we collect all posts that mention ``china'' or ``chinese'' and apply the same pre-processing strategy as the section ``Racial Slurs'', which results in 2,193,410 posts from 4chan's /pol/ and 26,183,882 posts from Reddit.
Similar to~\cite{RJZBCSW21}, we consider posts with SEVERE\_TOXICITY score $\ge$ 0.8 as toxic posts and filter out 224,087 posts from 4chan's /pol/ and 388,060 posts from Reddit as the trainset of each Web community.
We elect SEVERE\_TOXICITY as it is less sensitive to more mild forms of toxicity, such as comments that include positive uses of curse words~\cite{SEVERETOXIC}.

We leverage Top2vec~\cite{A20} as our topic modeling method to extract topics from the post.
After training with its default setting, the models generate 924 topics for 4chan's /pol/ and 1,542 topics for Reddit, respectively.
We then hierarchically reduce the total number of topics to smaller values ranging from 10 to 80 with a stepping size of 10. 
Concretely, we use u\_mass~\cite{MWLM11} to evaluate the quality of the models with different numbers of topics where a higher u\_mass indicates a better model.
We calculate the u\_mass for each model and pick the one with the highest u\_mass value as our final topic model, which is 20 for both 4chan's /pol/ and Reddit.
Figure~\ref{figure: topic_word_cloud_pol} and~\ref{figure: topic_word_cloud_reddit} shows the word cloud of each topic for each web community. 
Note that we sort the topics by their popularity, e.g., topic 0 is the most popular one.

\mypara{Topics on 4chan's /pol/}
First, when looking at the results of 4chan's /pol/ (see Figure~\ref{figure: topic_word_cloud_pol}), we uncover some general topics including ethnics (topic 0), economics and commerce (topic 1), foreign relations (topic 4), China internal affairs (topic 6), Chinese government (topic 5), weapon and military (topic 15), as well as event-related topics, like COVID-19 (topic 3) and U.S. election (topic 7), depicting the panoramic view of diverse anti-China rhetoric.
For instance, the most popular topic on 4chan's /pol/ towards Chinese is topic 0, containing words referring to ethnics, mostly towards Chinese and other East Asians (e.g., ``asians'', ``koreans'', ``chinese'' as well as ``iq'', ``genes'').
A /pol/ user posted: \textit{``the chinese iq meme comes from the chinese government literally cherrypicking their best.  most of china and almost the entirety of southern asia are dumb.''}
Another /pol/ user posted: \textit{``most asians are awful in everyway, including the women. though japs, and occasionally korean or chinese, are decent like whites. the holocaust is a lie. whites are superior. most blacks are physically incompetent as much as they are mentally.''}

We also observe several topics containing multi-language words, like Chinese (topic 16), Portuguese (topic16), and German (topic18), which indicates the active participation of speakers using these languages in Sinophobic discussions.
For example, a /pol/ user who sets his country to Brazil posts 
\textit{``você é um preto socialista filho de uma puta que nunca se importou desse país inteiro estar se vendendo pra china.'' (in translation, ``you're a black socialist son of a bitch who never cared about this whole country selling to china.'')}
Interestingly, TikTok~\cite{TIKTOK}, a famous video-sharing application owned by the Chinese company, is frequently mentioned in toxic posts of /pol/ (topic 13) and is regarded as a weapon from China, e.g., \textit{`` tiktok is a chinese weapon designed to turn western children into weak pussies.''}, indicating the fear or hatred towards Chinese people.
These kinds of emotions and prejudice are also regarded as the origins of Sinophobia~\cite{B14}.

\mypara{Topics on Reddit}
For Reddit, we can observe a number of similar toxic topics as 4chan's /pol/, e.g.,  economics and commerce (topic 1), weapon and military (topic 8), internal affairs (topic 17),  ethnics (topic 3), foreign relations (topic 8), Chinese government (topic12), COVID-19 (topic 7), and U.S. election (topic 9).
However, toxic topics on Reddit are more diverse in language and cover a wider range of aspects than 4chan's /pol/.
For instance, the most popular topic on Reddit is a multi-language one, combining slurs in German, Italian, Spanish, and other languages.
Examples are \textit{``zusammen china ficken (in translation, fuck together china).''}, \textit{``la brutta china è realtà. (in translation, ugly china is reality.)''}, and \textit{``china filha da puta (in translation, china son of a bitch)}, respectively. 
``Yellow fever'', a strong sexual or romantic preference for persons of Asian descent~\cite{YELLOWFEVER}, is also one hot topic on Reddit (Rank top 6).
For example, a Reddit user posts: \textit{``i got a bitch up in china, i like to fuck her vagina.''}
In addition, we observe racial slurs in topics about daily life, including food (topic 10), game (topic 11), and stock (topic 18).
For example, \textit{``I ate chinese food so you better hope I don’t fart in your fucking face.''}, \textit{``stupid chinese cheaters messing up all the good battle royale games''}, \textit{``i feel like nio is the laughing stock of the trading world. lol fucking lying chinese hustle.''}

\mypara{Shift in topics}
Now, we discuss the shift of topics from the pre-pandemic period to the pandemic period (see Table~\ref{table:topic_per_4chan} and Table~\ref{table:topic_per_reddit}).
Overall, we find users' interests drastically shift to COVID-19 on both Web communities.
The shift ratio reaches $11\times$ on 4chan's /pol/ and $5.32\times$on Reddit.
We also observe that Chinese-goverment topics are the most toxic ones in the whole period on both /pol/ (ST = 0.86) and Reddit (ST = 0.85).
The scale of it exploded to $1.25\times$ on 4chan's /pol/ (topic 5) and $1.71\times$ on Reddit (topic 12) during the pandemic period, impling users express more toxic posts towards Chinese government with anger.

Meanwhile, we also notice the differences between the two Web communities.
For instance, the topic that blaming Biden as a Chinese spy is both on 4chan's /pol/ (topic 7) and Reddit (topic 9).
However, the discussion proportion increases $1.59\times$ in the pandemic period on /pol/ while it remains the similar ratio, i.e. $1.01\times$, on Reddit.
Besides, the most popular topic consisting of multi-language slurs on Reddit continues growing in the pandemic period, which implies that more countries join to contribute Sinophobic slurs.

\mypara{Takeaways}
In this section, we focus on the evolution of Sinophobic content. 
We analyze the magnitude of semantic change and find that ``china,'' ``chinese,'' and ``virus'' suffer higher changes than the baseline words like ``america,'' ``hk,'' and ``jew.''
For instance, compared to the pre-pandemic period on 4chan's /pol/, the average cosine similarity of ``china,'' ``chinese,'' and ``virus'' in the pandemic period drops 0.066, 0.065, and 0.076, respectively, while only 0.003, 0.057, -0.034 for ``america,'' ``hk,'' and ``jew,'' respectively.
To further understand the meaning of such changes, we visualize the diachronic word embeddings of words that are similar to ``china'' and ``chinese.''
Our observation reveals that the meaning of ``china'' and ``chinese'' shifts from the notation of the country to the pandemic words and Sinophobic slurs (see also Figure~\ref{figure: histwords}).
We then dive into the detailed Sinophobic topics and find that the most popular topic is related to Asian people on 4chan's /pol/, while on Reddit we find Sinophobic slurs shared in various languages.
Also, we find that both Web communities share some common Sinophobic topics, e.g., ethnics, weapon and military, etc.
However, compared to 4chan's /pol/, Reddit contains more Sinophobic topics related to daily life, including food (topic 10), game (topic 11), and stock (topic 18).
After the outbreaks of COVID-19, we find that users' interests drastically shift to COVID-19 on both Web communities.
The shift ratio reaches $11\times$ on 4chan's /pol/ (topic 3) and $5.32\times$ on Reddit (topic 7).
In the pandemic period, users also express more toxic posts towards the Chinese government, which is the most toxic topics on both fringe and mainstream Web communities.

\section{Conclusion}

In this paper, we investigate how online Sinophobia has evolved by performing a large-scale measurement from 2016 to 2021 over two Web communities (Reddit and 4chan's /pol/).
We first investigate the temporal pattern of the posts that related to China and analyze discovered Sinophobic slurs.
Our findings quantitatively reveal that Sinophobia was well established before COVID-19, most often sustained by directly or non-directly related political events.
While the COVID-19 pandemic greatly increased Sinophobia online, it also marked a sharp change in the \emph{kind} of Sinophobia exhibited: ``china'' and ``chinese'' shifted away from referring to the country/government of China towards Sinophobic slurs.
When exploring the characteristcs of Sinphobic topics across the Web communities we study, we find an overlap in Sinophobic topics like ethnicities, however, Reddit also has much more benign topics like food.
Much like Xingtian, although COVID-19 has made Sinophobia much more grotesque, it has been an ever present part of online discussion.

\section*{Acknowledgments}

This work is partially funded by the Helmholtz Association within the project ``Trustworthy Federated Data Analytics'' (TFDA) (funding number ZT-I-OO1 4) and supported by the National Science Foundation (grant number 2046590).

\balance
\bibliographystyle{plain}
\bibliography{normal_generated_py3,other}

\end{document}